\documentclass[preprint,aps,prc,showpacs,nofootinbib,secnumarabic,superscriptaddress]{revtex4-1}

\usepackage{graphicx}
\usepackage{bm}
\usepackage{color}

\newcommand{\ltsim}{\protect\raisebox{-0.5ex}{$\:\stackrel{\textstyle <}
	{\sim}\:$}}

\newcommand{\bvec}[1]{\ensuremath{\boldsymbol{#1}}}

\begin{document}

\title{Elastic scattering, polarization and absorption of relativistic antiprotons on nuclei}

\author{A.B. Larionov}
\affiliation{Institut f\"ur Theoretische Physik, Universit\"at Giessen,
  D-35392 Giessen, Germany}
\affiliation{National Research Center "Kurchatov Institute", 123182 Moscow, Russia}
\author{H. Lenske}
\affiliation{Institut f\"ur Theoretische Physik, Universit\"at Giessen,
             D-35392 Giessen, Germany}

\date{\today}

\begin{abstract}
We perform Glauber model calculations of the antiproton-nucleus elastic and quasielastic scattering
and absorption in the beam momentum range $\sim 0.5\div10$ GeV/c.   
A good agreement of our calculations
with available LEAR data and with earlier Glauber model studies of the $\bar p A$ elastic scattering
allows us to make predictions at the beam momenta of $\sim 10$ GeV/c, i.e. at the regime of the PANDA
experiment at FAIR. 
The comparison with the proton-nucleus
elastic scattering cross sections shows that the diffractive minima are much deeper in the $\bar p A$ case
due to smaller absolute value of the ratio of the real-to-imaginary part of the elementary elastic amplitude.
Significant polarization signal for $\bar p A$ elastic scattering at 10 GeV/c is expected.
We have also revealed a strong dependence of the $\bar p A$ absorption cross section on the slope parameter
of the transverse momentum dependence of the elementary $\bar pN$ amplitude. The $\bar p A$ optical potential
is discussed.
\end{abstract}


\maketitle

\section{Introduction}
\label{intro}

The Glauber model (GM) is an extremely successful theoretical method to describe 
exclusive and semi-exclusive interactions of the moderately relativistic (beam momentum $p_{\rm lab} \sim 1\div10$ GeV/c) particles with nuclei.  
Since its formulation \cite{Glauber} the GM has been widely used to describe interactions of protons with nuclei, in-particular,
for testing nuclear structure by elastic and quasielastic $pA$ scattering \cite{Czyz:1967vnq,Bassel:1969gs,Glauber:1970jm}.
Once the elementary amplitude is known, as it is largely the case for the $pN$ scattering, one can apply the GM for the studies
of the nucleon and neutron density profiles. In late 70-s this has been in the focus of the analyses of the elastic $pA$ scattering
experiments at 1 GeV in Gatchina \cite{Alkhazov:1975vp} and Saclay \cite{Alkhazov:1977ecl,Chaumeaux:1977sf}
(see ref. \cite{Alkhazov:1978et} for the review of the GM studies of $pA$ scattering).

The bulk of phenomenological information on the proton and neutron density distributions in $\beta$-stable nuclei is nowadays already
quite rich. This allows one to use the GM with fixed proton and neutron density profiles for the studies of the unknown hadron-nucleon
($hN$) elastic amplitudes.
Of special interest is the ratio of the real-to-imaginary part of the forward scattering amplitude,
\begin{equation}
   \alpha = {\mbox{Re}f_\tau(0) \over \mbox{Im}f_\tau(0)}~,   \label{alpha}
\end{equation}
where $\tau=+1/2 (-1/2)$ denotes the isospin of the target proton (neutron)\footnote{Where it can not lead to misunderstanding,
we will also use literal notations $\tau=p,n$.}.
Using complex nuclear targets is, in particular, the only way to determine the sign and the value of $\alpha$ of the hadron-neutron
amplitude\footnote{See, for example, ref. \cite{Mahalanabis:1992ee} where $\alpha$ of the $\bar p n$ amplitude was extracted
from $\bar p d$ elastic scattering. In contrast, in the case of the $pp$ and $\bar p p$ elastic scattering the interference with the Coulomb
amplitude at small scattering angles can be used for the determination of $\alpha$, cf. ref. \cite{Jenni:1977kv}.}.  

The aim of this work is twofold: (i) Based on the phenomenological parameters of the $\bar p N$ elastic amplitude to provide the GM predictions
on the $\bar p A$ differential elastic scattering cross sections including polarization, and on $\bar p A$ absorption cross sections
in the regime of FAIR ($p_{\rm lab} \sim 10$ GeV/c).
(ii) To study the sensitivity of the results to the choice of the parameters of the $\bar p N$ elastic amplitude. 

There are several interesting features of the $\bar p A$ elastic scattering.
In contrast to the $pp$ amplitude, the $\bar p p$ amplitude is forward-peaked at small beam momenta
(see Fig.~\ref{fig:slopes} below).
This explains why the GM analysis \cite{Dalkarov:1984lcx,Dalkarov:1986tu} of the LEAR data on $\bar p A$ elastic scattering
has been very successful even at the beam energy as low as 46.8 MeV ($p_{\rm lab}=300$ MeV/c). 
Since the total $\bar p N$ cross section, 
$\sigma_{\bar pN}$, is large due to annihilation, the $\bar p A$ scattering is largely dominated by nuclear interactions, 
except for the extreme forward angles and diffractive minima where the Coulomb interaction provides the main contribution 
\cite{Dalkarov:1984lcx,Dalkarov:1986tu}.
Thus, the $\bar p A$ scattering can be used to get information on the $\bar p N$ elastic amplitude.
At the beam momenta below 1 GeV/c, the experimental determination of $\alpha$ of the $\bar p N$ amplitude
makes possible to test how well $G$-parity works by calculating the scattering amplitude from optical
$\bar N N$ potentials (e.g. Bonn \cite{Hippchen:1987jt}, Paris \cite{Cote:1982gr,ElBennich:2008vk},
chiral \cite{Kang:2013uia}). The information on the $\bar p$ polarization 
in $\bar p A$ scattering would be useful to constrain the spin-orbit parts of $\bar N N$ potentials. 
At higher beam momenta, the information on the $\bar p N$ amplitude could constrain parameters of the Regge model 
\cite{Collins}. 

Another quantity of interest is the $\bar p A$ absorption cross section, $\sigma_{\rm abs}$.
This observable not only tests the mechanisms of $\bar p$ interaction with nuclei,
but also serves as an input in the cosmic antiproton flux calculations (cf. \cite{Moskalenko:2001ya}).
The latter are important for the analyses of the PAMELA
(Payload for Antimatter Matter Exploration and Light-nuclei Astrophysics)  \cite{Hooper:2014ysa}
and preliminary AMS-02 (Alpha Magnetic Spectrometer) \cite{Chen:2015kla} data on the $\bar p/p$ ratio,
where the possibility of the dark matter annihilation is currently under discussions.

Antiproton annihilation may also be a tool for the radiotherapy. The $\bar p$ annihilation
leads to much higher dose deposition within the Bragg peak in comparison to the proton irradiation
\cite{Holzscheiter:2006ya}. Modeling antiproton interactions with ordinary media requires 
realistic $\bar p$ absorption cross sections on nuclei at low energies. 

The $\bar p$ absorption is surface-dominated ($\sigma_{\rm abs} \propto A^{2/3}$,  cf. Fig.~2 in ref. \cite{Balestra:1985kn})
due to large $\sigma_{\bar pN}$ and, thus, is strongly sensitive to the radius of the nucleus.
It has been demonstrated in ref. \cite{Lenske:2005nt} on the basis of the optical model that the $\bar p A$ absorption
allows to test the radii of unstable isotopes.
On the other hand, as has been shown in ref. \cite{Larionov:2009tc} on the basis of the GiBUU model with relativistic mean fields,
the attractive $\bar p A$ interaction at low beam momenta ($p_{\rm lab} \ltsim 1$ GeV/c)
bends the antiproton trajectory towards the nucleus, thus, tending to increase the absorptive interactions.
In the recent works \cite{Lee:2013rxa,Lee:2015hma}, the GM extended by attractive Coulomb and nuclear potentials has been 
formulated and successfully applied to calculate the $\bar p A$ and $\bar n A$ annihilation cross sections
at 0.1 GeV/c $\ltsim p_{\rm lab} \ltsim$ 3 GeV/c. 

The structure of our work is as follows. In section \ref{basic} we give a summary of the Glauber 
multiple scattering formalism. 
Section \ref{NumRes} addresses the comparison of the GM results with LEAR data on differential cross sections of $\bar p A$ elastic
scattering and polarization and BNL data on differential cross sections of $p A$ quasielastic scattering.
Predictions for $\bar p A$ elastic and quasielastic scattering and polarization at FAIR regime
($p_{\rm lab}=10$ GeV/c) are given.
In section \ref{NumRes}, also the absorption of antiprotons on nuclei is considered.
We demonstrate that at 0.5 GeV/c $\ltsim p_{\rm lab} \ltsim$ 10 GeV/c
the absorption cross section is well described within the GM if one takes into account the momentum-transfer dependence
of the $\bar p N$ scattering amplitude and study the sensitivity of our results to the choice
of the slope parameter of the transverse momentum dependence. 
Finally, sec. \ref{concl} summarizes the main results of our work. 
Appendix \ref{SPdensities} contains the explicit formulas for the nucleon densities in the case of shell model
wave functions.

\section{Glauber model}
\label{basic}

According to the GM \cite{Glauber,Glauber:1970jm} the scattering amplitude on the nucleus summed over all possible
nucleons-scatterers is expressed as
\begin{eqnarray}
    F_{fi}(\bm{q}) &=& \frac{ik}{2\pi} \int d^2 b\, e^{-i\bm{q} \cdot \bm{b}} \int d x_1 \ldots dx_A\,
           \psi_{f}^*(x_1,\ldots,x_A)
           \psi_{i}(x_1,\ldots,x_A)  \\
    && \times \left(1 - \prod_{j=1}^A[1-\Gamma_{T_j}(\bm{b}-\bm{r}_{jt})]\right)~,     \label{F_fi}
\end{eqnarray}
where $\bm{q}=\bm{k}^\prime-\bm{k}$ is the momentum transfer. $\bm{k}$ and $\bm{k}^\prime$ are,
respectively, the incoming and outgoing particle momenta in the laboratory frame.
$\psi_{i}$ and $\psi_{f}$ are the wave functions of the initial ($i$) and final ($f$) nuclei. $x_j \equiv (\bm{r}_j,\lambda_j,T_j)$
denotes the $j$-th coordinate, $\bm{r}_j$, spin, $\lambda_j=\pm 1/2$, and isospin, $T_j=\pm 1/2$, variables of the wave function.
The integration over $dx_j$ is a short-hand notation for the integration over $d^3r_j$ and summation over $\lambda_j$ and $T_j$.
The profile functions of nucleons are defined as the Fourier transforms of the elementary elastic 
$hN$ scattering amplitudes, $f_\tau(\bm{q}_{t})$, in the transverse momentum transfer, $\bm{q}_{t}$:
\begin{equation}
   \Gamma_\tau(\bm{b}) = \frac{1}{2 \pi i k}  
               \int d^2q_{t}\, e^{i\bm{q}_{t} \cdot \bm{b}} f_\tau(\bm{q}_{t})~.   \label{Gamma_tau}
\end{equation}

For heavy nuclei the derivation of the GM amplitude of the particle-nucleus interaction on the basis of the Feynman diagrams has been given
by Gribov \cite{Gribov:1968gs} and Bertocchi \cite{Bertocchi:1972cj}. In the GM amplitude the elementary $hN$ amplitudes are set on-shell.
Thus they can be expressed as functions of the transverse momentum transfers only. If an on-shell $hN$ amplitude quickly drops
with transverse momentum transfer, then one can neglect the transverse momentum transfers in the propagators of the scattered
hadron $h$ which allows to factorize them out in the partial amplitudes (with fixed set of nucleons-scatterers).
This corresponds to the eikonal approximation, i.e. to disregarding the influence of the multiple momentum
transfers on the particle trajectory. Hence it is clear that the GM should better work when elementary amplitudes
are forward peaked in the $hN$ center-of-mass (c.m.) frame, i.e. dominated by $t$-channel interactions.

Generally, the wave functions of the initial and final nuclei should contain the Pauli exclusion effect.
In the shell model, the Pauli exclusion is treated by using the Slater determinant of the occupied single-particle states:
\begin{equation}
  \psi_{i}(x_1,\ldots,x_A) = \frac{1}{\sqrt{A!}}
    \left|
    \begin{array}{llll}
      \phi_1(x_1) & \phi_1(x_2) & \ldots & \phi_1(x_A) \\
      \phi_2(x_1) & \phi_2(x_2) & \ldots & \phi_2(x_A) \\
      \ldots \\
      \ldots \\
      \phi_A(x_1) & \phi_A(x_2) & \ldots & \phi_A(x_A)
    \end{array}
    \right| \equiv  \frac{1}{\sqrt{A!}} ||\phi_m(x_n)||~,                              \label{psi_i}
\end{equation}
and similar for the wave function of the final nucleus,
\begin{equation}
  \psi_{f}(x_1,\ldots,x_A) = \frac{1}{\sqrt{A!}} ||\varphi_m(x_n)||~,    \label{psi_f}
\end{equation}
where $\phi_m(x)$ and $\varphi_m(x)$ are the single-nucleon wave functions in the initial and final nuclei,
respectively. The index $m$ enumerates the occupied single-particle states. 
The single-nucleon wave functions of the initial nucleus constitute the orthogonal set (see also Eq.(\ref{Glob_orhthog_expl})
of Appendix~\ref{SPdensities}), i.e.
\begin{equation}
  \int dx \phi_{m^\prime}^*(x) \phi_m(x) = \delta_{m^\prime m}~,   \label{Glob_orthog}
\end{equation}  
and similar for those of the final nucleus.

It is convenient, following ref. \cite{Bassel:1969gs}, to introduce notation
\begin{equation}
  O_j \equiv 1-\Gamma_{T_j}(\bm{b}-\bm{r}_{jt})~.     \label{O_j}
\end{equation}
Then, since the product $\prod_{j=1}^A O_j$ is symmetric with respect to the simultaneous interchange of the isospin
and coordinate variables, we can transform its matrix element as follows:
\begin{eqnarray}
  && \langle \psi_{f}|\prod_{j=1}^A O_j|\psi_{i}\rangle
  \equiv \int d x_1 \ldots dx_A\,\psi_{f}^*(x_1,\ldots,x_A) \prod_{j=1}^A O_j \psi_{i}(x_1,\ldots,x_A)  \nonumber \\
  && = \int d x_1 \ldots dx_A\, \prod_{j=1}^A \varphi_j^*(x_j) O_j ||\phi_m(x_n)||
     = ||\int d x_n \varphi_n^*(x_n) O_n \phi_m(x_n)||    \nonumber \\
  && = ||\langle \varphi_n|\phi_m\rangle -  \int d x \varphi_n^*(x) \Gamma_{T}(\bm{b}-\bm{r}_{t}) \phi_m(x)||~,  \label{prod_O_j}
\end{eqnarray}
where in the last step we replaced $x_n \to x$ as the integrals over $dx_n$ inside the antisymmetrized product can be taken
independently.
In the case of elastic scattering, $\psi_i=\psi_f$ and the wave functions of the occupied states in the initial and final nuclei are identical,
i.e. $\varphi_n(x) = \phi_n(x)$.
Thus, disregarding the nondiagonal transitions Eq.(\ref{prod_O_j}) can be approximately rewritten as
\begin{eqnarray}
  \langle \psi_{i}|\prod_{j=1}^A O_j|\psi_{i}\rangle & \simeq &
  \prod_{j=1}^A (1 - \int d x \Gamma_{T}(\bm{b}-\bm{r}_t) |\phi_j(x)|^2) \nonumber \\
  &=& \prod_{j=1}^A (1 - \int d^3r \sum\limits_{\lambda,T}\Gamma_{T}(\bm{b}-\bm{r}_t)|\phi_j(\bm{r},\lambda,T)|^2)~.   \label{IPM}
\end{eqnarray}
(This expression can be also directly obtained by using the independent particle model, i.e. expressing the wave function of the nucleus
as a product of the single-particle wave functions instead of the Slater determinant.)
Introducing the single-particle densities (see Eq.(\ref{rho_j_expl}) of Appendix~\ref{SPdensities} for the explicit formula),
\begin{equation}
   \rho_j(\bm{r})=\sum\limits_{\lambda,T} |\phi_j(\bm{r},\lambda,T)|^2~,   \label{rho_j}
\end{equation}
with the normalization condition
\begin{equation}
   \int \rho_j(\bm{r}) d^3 r =1~,    \label{NormCond}
\end{equation}
and using Eq.(\ref{IPM}) the elastic amplitude can be now expressed as
\begin{equation}
   F_{\rm el}(\bm{q}) \equiv F_{ii}(\bm{q}) = \frac{ik}{2\pi} \int d^2 b\, e^{-i\bm{q} \cdot \bm{b}}
    \left(1 - \prod_{j=1}^A[1-\int d^3r \rho_j(\bm{r})\Gamma_{\tau_j}(\bm{b}-\bm{r}_t)]\right)~.     \label{F_el}
\end{equation}

It has been shown in the previous studies \cite{Bassel:1969gs,Alkhazov:1978et} that the full treatment
of the Pauli exclusion principle results in the corrections to the angular differential cross section of 1 GeV proton
scattering by $\sim 5\div10\%$ with respect to the use of the independent particle model Eq.(\ref{F_el}).
Corrections of the same order are provided by short-range correlations which are the consequence
of the hard-core repulsion of nucleons at short distances $\ltsim 1$ fm
(cf. Fig.~14 in ref. \cite{Alkhazov:1978et}). The study of these effects goes beyond the scope of our work.  

For heavy nuclei, $A \gg 1$, the product in Eq.(\ref{F_el}) can be replaced by the exponent:
\begin{equation}
   F_{\rm el}(\bm{q}) \simeq \frac{ik}{2\pi} \int d^2 b\, e^{-i\bm{q} \cdot \bm{b}}
    \left(1 - e^{i\chi_N(b)}\right)=ik \int\limits_0^\infty db b J_0(qb) \left(1 - e^{i\chi_N(b)}\right)~,   \label{F_el_largeA}
\end{equation}
with
\begin{equation}
   \chi_N(\bm{b}) = i \sum_{j=1}^A \int d^3r \rho_j(\bm{r}) \Gamma_{\tau_j}(\bm{b}-\bm{r}_t)
             = \frac{1}{2\pi k} \sum_{\tau=\pm1/2} \int d^2 q_{t}\, e^{i\bm{q}_{t} \cdot \bm{b}}
                 S_\tau(-\bm{q}_{t}) f_\tau(\bm{q}_{t})     \label{chi_N}
\end{equation}
being the nuclear phase-shift function. The form factors of the proton and neutron densities are defined as
\begin{equation}
   S_\tau(\bm{q}) = \int d^3 r e^{i \bm{q} \cdot \bm{r}} \rho_\tau(\bm{r})~,   \label{S_tau}
\end{equation}
with
\begin{equation}
   \rho_\tau(\bm{r})= \sum_{j=1}^A \rho_j(\bm{r}) \delta_{\tau,\tau_j}~.   \label{rho_tau}
\end{equation}

Note that Eq.(\ref{F_el_largeA}) can be also obtained by applying the eikonal approximation to the Schr\"odinger equation 
for scattering in the external potential 
\begin{equation}
   U(\bm{r}) = -\frac{1}{(2\pi)^2 \sqrt{m_p^2+k^2}} \sum_\tau \int d^3 q  S_\tau(-\bm{q})
                                f_\tau(\bm{q}) e^{i\bm{q} \cdot \bm{r}}~,     \label{U}
\end{equation}
where $m_p$ is the projectile mass.
In this case the phase-shift function is expressed as an integral of the potential along straight-line particle trajectory
(cf. \cite{QM})
\begin{equation}
   \chi_N(\bm{b}) =  -\frac{1}{v}\int\limits_{-\infty}^{+\infty} dz\, U(\bm{b},z)~,   \label{chi_N_U}
\end{equation}
where $v=k/\sqrt{m_p^2+k^2}$ is the projectile velocity.

\subsection{Coulomb correction}
\label{Coulomb}

The relation (\ref{chi_N_U}) can be used to accommodate the Coulomb effects in the GM.
According to ref. \cite{Glauber:1970jm} this is reached by replacing $\chi_N(b) \to \chi_N(b) + \chi_C(b)$
in Eq.(\ref{F_el_largeA}). The Coulomb phase-shift function is 
\begin{equation}
   \chi_C(\bm{b}) = -\frac{1}{v} \int\limits_{-R_{\rm scr}}^{+R_{\rm scr}} dz\,
                \int d^3r^\prime\, U_C(|\bm{r}-\bm{r}^\prime|) \rho_{\rm ch}(\bm{r}^\prime)~,    \label{chi_C}
\end{equation}
where $U_C(r)=\pm e^2/r$ is the Coulomb potential acting between the scattered proton (+) or antiproton
(-) and the target proton, $\rho_{\rm ch}(\bm{r})$ is the charge density normalized to $Z$, 
$R_{\rm scr}$ is the screening radius introduced in order to ensure the convergence of the integral.
In the limit $R_{\rm scr} \gg b$ it is possible to derive the semi-analytic form of Eq.(\ref{chi_C})
\cite{Alkhazov:1978et,Dalkarov:1984lcx} \footnote{
In the original paper of Glauber and Matthiae \cite{Glauber:1970jm} the condition $R_{\rm scr} \gg q^{-1}$ is quoted. 
However, the presence of the Bessel function $J_0(qb)$ in Eq.(\ref{F_el_largeA}) which decreases as $\propto 1/\sqrt{qb}$
with increasing $qb$ ensures that $b \sim q^{-1}$.}:
\begin{eqnarray}
   \chi_C(\bm{b}) &=& \chi_{\rm scr} + \chi_0(b) + \chi_1(b)~,
   ~~~\chi_{\rm scr} = -2\xi\log(2R_{\rm scr}k)~,
   ~~~\chi_0(b) = 2\xi\log(kb)~,  \nonumber     \\
   \chi_1(b) &=& 8\pi\xi\int\limits_b^\infty dr r^2 \tilde\rho_{\rm ch}(r)
                \left(\log\left(\frac{1+\sqrt{1-(b/r)^2}}{b/r}\right)
                      -\sqrt{1-(b/r)^2}\right)      \label{chi_C_simpl}
\end{eqnarray}
with
\begin{equation}
  \xi = \pm \frac{Ze^2}{v}    \label{xi}
\end{equation}
and $\tilde\rho_{\rm ch}(r) = Z^{-1}\rho_{\rm ch}(r)$.
Using Eqs.(\ref{chi_C_simpl}) the elastic amplitude on the nucleus with Coulomb correction
can be written as
\begin{equation}
    e^{-i\chi_{\rm scr}} F_{\rm el}(\bm{q}) =
    F_C(q) +  \frac{ik}{2\pi} \int d^2b\, e^{-i\bm{q} \cdot \bm{b} + i\chi_0(b)}
              \left(1 - e^{i[\chi_N(b)+\chi_1(b)]}\right)~,    \label{F_el_Coul_corr}
\end{equation}
where $F_C(q)$ is the Coulomb scattering amplitude on a point-like nucleus:
\begin{equation}
     F_C(q) =-\frac{2 \xi k}{q^2} e^{i\phi_C}~,
    ~~~\phi_C=-2\xi\log\left(\frac{q}{2k}\right)+2\eta~,
    ~~~\eta = \arg\Gamma(1+i\xi)~.                         \label{F_C}
\end{equation}
As we see, the screening radius introduces the ambiguity in the phase of the full amplitude only and
thus does not influence any observables.

\subsection{Spin effects}
\label{Spin}

The elementary $\bar p p$ and $\bar p n$ amplitudes include the nuclear central
and spin-orbit interactions \cite{Glauber:1978fw,Osland:1978vu,Tan:1989gd}:
\begin{equation}
   f_\tau(\bm{q})=A_\tau(q)
   +C(q)\bvec{\sigma} \cdot [\bm{q} \times \hat{\bm{k}}]~,~~~\hat{\bm{k}}=\bm{k}/k~,      \label{f_tau}
\end{equation}
where 
\begin{eqnarray}
    A_\tau(q) &=& \frac{i k}{4\pi} \sigma_\tau (1-i\alpha_\tau) e^{-\beta_\tau q^2/2}~,   \label{A_tau}   \\
    C(q) &=& \frac{i k}{4\pi} \sigma_c (1-i\alpha_c) e^{-\beta_c q^2/2}~.        \label{C}
\end{eqnarray}
In Eq.(\ref{A_tau}), $\sigma_\tau$, $\alpha_\tau$ and $\beta_\tau$ are, respectively, the total cross section, the ratio of the real-to-imaginary part
and the slope of momentum transfer dependence of $\bar p p$ ($\tau=p$) and $\bar p n$ ($\tau=n$) central amplitudes.
For simplicity, we have chosen a similar analytic form for the spin-orbit amplitude and neglected the isospin dependence of its parameters,
$\sigma_c, \alpha_c$ and $\beta_c$.

The nuclear phase-shift function includes the spin-independent and spin-dependent components:
\begin{eqnarray}
   \chi_N(\bm{b}) &=& \chi_N^{(0)}(b) + \chi_N^{(S)}(b)~, \\
   \chi_N^{(0)}(\bm{b}) &=& \frac{1}{2\pi k} \sum_\tau \int d^2 q_{t}\, e^{i\bm{q}_{t} \cdot \bm{b}}
                       S_\tau(-\bm{q}_{t}) A_\tau(q_t)~,  \label{chi_N^0}    \\
   \chi_N^{(S)}(\bm{b}) &=& \frac{1}{2\pi k} \sum_\tau \int d^2 q_{t}\, e^{i\bm{q}_{t} \cdot \bm{b}}
                       S_\tau(-\bm{q}_{t})
                       C(q_t)\bvec{\sigma} \cdot [\bm{q}_{t} \times \hat{\bm{k}}]~.   \label{chi_N^S}
\end{eqnarray}
The spin-dependent component can be rewritten in a factorized form as follows:
\begin{eqnarray}
   \chi_N^{(S)}(\bm{b}) &=& \bvec{\sigma} \cdot [\hat{\bm{k}}\times\hat{\bm{b}}] \psi(b)~,
   ~~~\hat{\bm{b}}=\bm{b}/b~,                             \label{chi_N^S_fact}\\
   \psi(b) &=& \frac{i}{2\pi k} \frac{d}{db}   
               \sum_\tau \int d^2 q_{t}\, e^{i\bm{q}_{t} \cdot \bm{b}}
                       S_\tau(-\bm{q}_{t}) C(q_t)~.         \label{psi}
\end{eqnarray}
By using the identity
\begin{equation}
   e^{i\chi_N^{(S)}(\bm{b})} = \cos\psi(b) 
   + i \bvec{\sigma} \cdot [\hat{\bm{k}}\times\hat{\bm{b}}] \sin\psi(b)  \label{ident}
\end{equation}
we can now split the total elastic amplitude on the nucleus to the spin-independent and spin-dependent
components:
\begin{eqnarray}
    F_{\rm el}(\bm{q}) &=& F(\bm{q}) + F^{(S)}(\bm{q})~, \\
    F(\bm{q}) &=& 
    F_C(q) +  \frac{ik}{2\pi} \int d^2b\, e^{-i\bm{q} \cdot \bm{b} + i\chi_0(b)}
              \left(1 - e^{i[\chi_N^{(0)}(b)+\chi_1(b)]} \cos\psi(b) \right)~,           \label{F}  \\
    F^{(S)}(\bm{q}) &=&
    -\frac{ik}{2\pi} \int d^2b\, e^{-i\bm{q} \cdot \bm{b} + i[\chi_N^{(0)}(b)+\chi_0(b)+\chi_1(b)]}     
      i\bvec{\sigma} \cdot [\hat{\bm{k}}\times\hat{\bm{b}}] \sin\psi(b)~,     \label{F^S}
\end{eqnarray}
where the overall screening phase of Coulomb interaction is set to zero (cf. Eq.(\ref{F_el_Coul_corr})).
The spin-dependent amplitude, Eq.(\ref{F^S}), can be rewritten in a factorized form:
\begin{eqnarray}
    F^{(S)}(\bm{q}) &=& G(q) \bvec{\sigma} \cdot [\hat{\bm{q}}\times\hat{\bm{k}}]~,  \label{F^S_fact}   \\
    G(q) &=& -\frac{ik}{2\pi} \frac{d}{dq} \int \frac{d^2b}{b}\,  
    e^{-i\bm{q} \cdot \bm{b} + i[\chi_N^{(0)}(b)+\chi_0(b)+\chi_1(b)]} \sin\psi(b) \\
         &=& ik \int\limits_0^\infty db b J_1(qb) e^{i[\chi_N^{(0)}(b)+\chi_0(b)+\chi_1(b)]} \sin\psi(b)~.  \label{G}
\end{eqnarray}

Thus, the elastic amplitude on the nucleus can be finally expressed as
\begin{equation}
    F_{\rm el}(\bm{q}) = F(q) + G(q) \bvec{\sigma} \cdot \hat{\bm{n}}~,        \label{F_el_final}
\end{equation}
where $\hat{\bm{n}}=[\hat{\bm{q}}\times\hat{\bm{k}}]$ is the normal vector to the scattering plane.
The differential scattering cross sections for the antiproton spin parallel ($\uparrow$) and antiparallel ($\downarrow$)
to $\hat{\bm{n}}$ are different:
\begin{equation}
      \frac{d \sigma_{\uparrow}}{d\Omega_{\rm lab}} = |F(q)+G(q)|^2~,
   ~~~\frac{d \sigma_{\downarrow}}{d\Omega_{\rm lab}} = |F(q)-G(q)|^2~.             \label{dsigmadOmega_pol}
\end{equation}
This leads to the nonvanishing polarization (analysing power) $P$ defined as the average of the Pauli $\bvec{\sigma}$ 
projection on $\hat{\bm{n}}$  \cite{Hoshizaki:1968wi,Osland:1978vu,Tan:1989gd}:
\begin{equation}
   P=\langle\bvec{\sigma} \cdot \hat{\bm{n}}\rangle 
      = {\frac{d \sigma_{\uparrow}}{d\Omega_{\rm lab}}
          -\frac{d \sigma_{\downarrow}}{d\Omega_{\rm lab}}
          \over
       2 \frac{d \sigma_{\rm el}}{d\Omega_{\rm lab}} }
    =\frac{2\mbox{Re}[F(q)G^*(q)]}{|F(q)|^2 + |G(q)|^2}~,    \label{P}
\end{equation}
where $d\sigma_{\rm el}/d\Omega_{\rm lab}$ is the spin-averaged elastic cross section:  
\begin{equation}
   \frac{d \sigma_{\rm el}}{d\Omega_{\rm lab}}
   =\frac{1}{2}\left(\frac{d \sigma_{\uparrow}}{d\Omega_{\rm lab}}
                     +\frac{d \sigma_{\downarrow}}{d\Omega_{\rm lab}}
               \right)
   = |F(q)|^2 + |G(q)|^2~.           \label{dsigmadOmega}
\end{equation}

The differential cross section in the c.m. frame of the colliding system is expressed as
\begin{equation}
   \frac{d \sigma_{\rm el}}{d\Omega_{\rm c.m.}} = \left(\frac{k_{c.m.}}{k}\right)^2 (|F(q)|^2 + |G(q)|^2)~,   \label{dsigmadOmega_cm}
\end{equation}
(and similar for the cross sections with polarized projectile), 
where $k_{c.m.}=kM_A/\sqrt{s}$ is the projectile momentum in the c.m. frame.
Here, $M_A$ is the mass of the nuclear target, and $\sqrt{s}$ is the total energy in the c.m. frame
which is given by $s=(E_p+M_A)^2-\bm{k}^2=m_p^2+M_A^2+2E_pM_A$ with $E_p=\sqrt{m_p^2+\bm{k}^2}$.
Equation (\ref{dsigmadOmega_cm}) can be simply obtained by expressing the transverse momentum transfer
in the laboratory and c.m. frames,  $k\sin\Theta_{\rm lab} \simeq k_{c.m.} \sin\Theta_{c.m.}$.

For further reference we also give a formula for the polarization in $\bar p p$ elastic scattering described
by the amplitude (\ref{f_tau}):
\begin{equation}
   P=\frac{2\mbox{Re}[A_p(q)C^*(q)q_t]}{|A_p(q)|^2 + 2 |C(q)|^2 q_t^2}~.    \label{P_pbarp}
\end{equation}
A factor of two in the denominator is due to taking into account the proton spin, $\bvec{\sigma}_p/2$.
This is reached by the replacement $\bvec{\sigma} \to \bvec{\sigma} + \bvec{\sigma}_p$ in Eq.(\ref{f_tau})
(see also discussion after Fig.~\ref{fig:pol_pbarp} in subsec. \ref{results_El}).
   
\subsection{Center-of-mass correction}
\label{CMcorr}

The shell model wave function, Eq.(\ref{psi_i}), does not correspond to the true target nucleus at rest
because its c.m. motion is not yet projected out. Moreover, the recoil momentum is not included in
the shell model wave function of the final nucleus, Eq.(\ref{psi_f}). 
The correct wave functions of the initial and final nuclei ($\nu=i,f$),
\begin{equation}
  \psi_\nu(x_1,\ldots,x_A) = \frac{e^{i\bm{k}_\nu \cdot \bm{R}}}{\sqrt{V}} J^{-1/2} \tilde\psi_\nu(x_1^\prime,\ldots,x_A^\prime)~,    \label{psi_nu}
\end{equation}
contain the plane wave factor for the c.m. motion with a normalization volume $V$, 
and the correlated part, $\tilde\psi_\nu(x_1^\prime,\ldots,x_A^\prime)$, with $x_j^\prime \equiv (\bm{r}_j^\prime,\lambda_j,T_j)$
expressed in terms of the relative positions of the nucleons
\begin{equation}
   \bm{r}_j^\prime = \bm{r}_j - \bm{R}~, \label{r_j^prime}
\end{equation}
with respect to the nuclear c.m. position,
\begin{equation}
   \bm{R}=\frac{1}{A} \sum\limits_{j=1}^A \bm{r}_j~.   \label{R}
\end{equation}
$J=A^3$ is the Jacobian of coordinate transformation, cf. ref. \cite{Alkhazov:1978et}.

If one calculates the amplitude of Eq.(\ref{F_fi}) with wave functions (\ref{psi_nu}) 
then the following formula can be derived (cf. \cite{Glauber:1970jm,Alkhazov:1978et}):
\begin{eqnarray}
   F_{fi}(\bm{q}) &=& \frac{(2\pi)^3}{V} \delta^{(3)}(\bm{k}_i+\bm{k}-\bm{k}_f-\bm{k}^\prime)
            \tilde F_{fi}(\bm{q})~,    \label{F_fi_with_CM} \\
   \tilde F_{fi}(\bm{q}) &=& \frac{ik}{2\pi} \int d^2 b\, e^{-i\bm{q} \cdot \bm{b}} \int dx_1^\prime \ldots dx_A^\prime
           \delta^{(3)}(\sum_{j=1}^A\bm{r}_j^\prime)
           \tilde\psi_{f}^*(x_1^\prime,\ldots,x_A^\prime)
           \tilde\psi_{i}(x_1^\prime,\ldots,x_A^\prime)  \\
    && \times \left(1 - \prod_{j=1}^A[1-\Gamma_{T_j}(\bm{b}-\bm{r}_{jt}^\prime)]\right)~.   \label{tildeF_fi}
\end{eqnarray}
The amplitude $\tilde F_{fi}(\bm{q})$ is related to the scattering matrix as
\begin{equation}
   S_{fi} = \frac{(2\pi)^4\delta^{(4)}(k_i+k-k_f-k^\prime)}{(2k^0V 2{k^0}^\prime V)^{1/2} V} 
            4\pi i \tilde F_{fi}(\bm{q})~.   \label{S_fi}
\end{equation}
Thus, $\tilde F_{ii}(\bm{q})$ should replace $F_{\rm el}(\bm{q})$ in the above formulas (\ref{dsigmadOmega_pol})-(\ref{dsigmadOmega})
for the differential cross sections and polarization.

In the case of the harmonic oscillator (HO) shell model wave function, 
the spurious wave function of the c.m. motion can be exactly factorized-out \cite{Bethe37}. 
In this case it can be shown that
\begin{equation}
   \tilde F_{fi}(\bm{q}) = H_{\rm cm}(q) F_{fi}^{\rm SM}(\bm{q})~,    \label{cmCorr}
\end{equation}
where $F_{fi}^{\rm SM}(\bm{q})$ is calculated from Eq.(\ref{F_fi}) using the shell model with HO single-particle
wave functions. The correction factor is
\begin{equation}
    H_{\rm cm}(q) = e^{\langle R^2 \rangle q^2/6}~,     \label{H_cm}
\end{equation} 
with $\langle R^2 \rangle=6a_0^2/4A$, where $a_0$ is the parameter of the HO nuclear density profile, 
$\rho(r) \propto e^{-r^2/a_0^2}$.
For heavy nuclei, where the HO model is not applicable, there is no exact relation like (\ref{cmCorr}). 
However, we will use Eq.(\ref{cmCorr}) also for heavy nuclei by setting $\langle R^2 \rangle=\langle r^2 \rangle/A$ where
\begin{equation}
     \langle r^2 \rangle = \frac{1}{Z} \int d^3 r r^2 \rho_p(r)    \label{<r^2>}
\end{equation}
which is valid if one neglects spatial correlations of nucleons \cite{Alkhazov:1978et}.

\subsection{Quasielastic background}
\label{QE}

The projectile may rescatter incoherently on one or more nucleons, which will lead to
excitation of the nucleus. The differential cross section of such quasielastic scattering
is given by summing the partial cross sections for all possible states of the final nucleus
except the ground state:
\begin{equation}
   \frac{d \sigma_{\rm QE}}{d\Omega_{\rm lab}} = \sum_{f \neq i} |F_{fi}(q)|^2~.   \label{dsigma_QE}
\end{equation}
The sum in the right-hand side can be represented as the difference of the sum over all states of the final nucleus
(which is calculated by using the completeness relation) minus the elastic amplitude squared. 
Skipping the details of a rather lengthy derivation which can be found in ref. \cite{Glauber:1970jm}
we only give the final result:
\begin{eqnarray}
   \sum_{f \neq i} |F_{fi}(q)|^2
   &=& \left(\frac{k}{2\pi}\right)^2 \int d^2bd^2B \,
       e^{-i\bm{q} \cdot \bm{b} - \sum\limits_{\tau=n,p} \sigma_\tau T_\tau(B)}
       \left( e^{\Omega(\bm{b},\bm{B})} -1 \right) \nonumber \\
   && -\sum\limits_{\tau=n,p} |f_\tau(q)|^2\,
       \frac{1}{N_\tau} \left|\int d^2B\, e^{-i\bm{q} \cdot \bm{B} + i\chi_N(B)}\, T_\tau(B)\right|^2~, \label{QEbg}
\end{eqnarray} 
where $N_p = Z~,~N_n = A-Z$, and
\begin{eqnarray}
   \Omega(\bm{b},\bm{B}) &=& \sum\limits_{\tau=n,p} \int d^3r \rho_\tau(r) 
                   \Gamma_\tau(\bm{B}+\frac{\bm{b}}{2}-\bm{r}_t)
                   \Gamma_\tau^*(\bm{B}-\frac{\bm{b}}{2}-\bm{r}_t)   \nonumber \\
               &\simeq& \frac{1}{k^2}\sum\limits_{\tau=n,p} T_\tau(B)
                        \int d^2Q\, e^{i\bm{Q} \cdot \bm{b}} |f_\tau(Q)|^2~.    \label{Omega}
\end{eqnarray}
The transverse positions, $\bm{b}$ and $\bm{B}$, are, respectively, the difference and the half-sum of
the transverse positions which appear in the direct, $F_{fi}(q)$, and complex-conjugated, $F_{fi}^*(q)$, amplitudes.
The quantities
\begin{equation}
   T_\tau(B)=\int\limits_{-\infty}^{+\infty} dz \rho_\tau(B,z)      \label{T_tau} 
\end{equation}
are the thickness functions of the proton ($\tau=p$) and neutron ($\tau=n$) density. 

The function $\Omega(\bm{b},\bm{B})$ vanishes unless $b \ltsim 2a$
and $B \ltsim R$, where $a \equiv \beta_p^{1/2}$ is the interaction range and
$R$ is the nuclear radius. Thus, the variation of the transverse positions, $\bm{b}$
and $\bm{B}$, is governed by the short scale of hadronic interaction and long nuclear scale,
respectively.

The first term in Eq.(\ref{QEbg}) provides the main contribution to the QE scattering.
It is derived under condition of the large momentum transfer, $q \gg R^{-1}$.
One can decompose $\exp\{\Omega(\bm{b},\bm{B})\}$ in the powers of $\Omega(\bm{b},\bm{B})$
which is known as the multiple-scattering expansion \cite{Glauber:1970jm}. 
The first term of this expansion, i.e. the single-scattering term, has the following form:
\begin{equation}
  \frac{d \sigma_{\rm QE}^{(1)}}{d\Omega_{\rm lab}}
   = \sum\limits_{\tau=n,p} |f_\tau(q)|^2  \int d^2B e^{-\sum\limits_{\tau^\prime=n,p} \sigma_{\tau^\prime} T_{\tau^\prime}(B)}  T_\tau(B)~.   \label{dsigma_QE_1coll} 
\end{equation}
The second term in Eq.(\ref{QEbg}) is a correction to the single-scattering 
contribution, Eq.(\ref{dsigma_QE_1coll}), in the case of small momentum transfers, $q \sim R^{-1}$.
In numerical calculations we will apply the full expression, Eq.(\ref{QEbg}).

\subsection{Absorption of antiprotons on nuclei.}
\label{Absorp}

Experimentally, the absorption cross section is determined by the transmission technique 
(cf. ref. \cite{Cool:1970fn} for the principle of the total cross section determination and
 refs. \cite{Abrams:1972ab,Nakamura:1984xw} for the absorption cross section).
This essentially means extrapolating to zero polar angles, $\Theta \to 0$, the following expression:
\begin{equation}
   \sigma_{\rm abs}(\Theta) = \sigma_{\rm tot}(\Theta) 
                                     - \sigma_{\rm d}(\Theta)~,          \label{sigma_abs_Theta}
\end{equation}
where $\sigma_{\rm tot}(\Theta)$ is the partial cross section measured by the transmission counter
which covers the polar angles $\Theta_{\rm lab} <  \Theta$, i.e. the cross section of removing
the beam particle from the solid angle subtended by the transmission counter,
and $\sigma_{\rm d}(\Theta)$ is the integrated elastic diffraction cross section to the solid angle region
beyond the transmission counter,
\begin{equation}
   \sigma_{\rm d}(\Theta) = \int\limits_{\Theta_{\rm lab} > \Theta} d\Omega_{\rm lab} 
                           \frac{d \sigma_{\rm el}}{d\Omega_{\rm lab}}~. \label{sigma_d_Theta}
\end{equation}
Thus, the absorptive contribution to the total cross section is provided by all possible inelastic processes.
The Coulomb interaction makes the total cross section divergent at $\Theta \to 0$. 
However, this divergence is subtracted since it is entirely contained in the diffractive cross section.
As a result, the r.h.s of Eq.(\ref{sigma_abs_Theta}) saturates at moderate scattering angles, where
one can neglect the Coulomb interaction.

Thus, we will calculate the absorption cross section entirely disregarding Coulomb effects.
In this case the total cross section can be calculated from the optical theorem as
\begin{equation}
   \sigma_{\rm tot}=\frac{4\pi}{k}\,\mbox{Im}\,F_{\rm el}(0)
   = 2 \int d^2 b\,\left(1 - \mbox{Re}\,e^{i\chi_N(b)}\right)~.      \label{sigma_tot}
\end{equation}
The angle-integrated diffraction cross section can be easily evaluated in the small-angle approximation:
\begin{equation}
  \sigma_{\rm d} = \int d\Omega_{\rm lab} \frac{d\sigma}{d\Omega_{\rm lab}}
   \simeq \frac{1}{k^2} \int d^2q_t |F_{\rm el}(q_t)|^2 
    = \int d^2 b\, |1 - e^{i\chi_N(b)}|^2~.                          \label{sigma_d}
\end{equation}
The absorption cross section is then obtained from the expression
\begin{equation}
   \sigma_{\rm abs} = \sigma_{\rm tot} - \sigma_{\rm d}
   = \int d^2 b\,\left(1 - e^{-2\,{\rm Im}\,\chi_N(b)}\right)~.      \label{sigma_abs}
\end{equation}      
Using Eq.(\ref{chi_N_U}) this expression can be also equivalently rewritten in terms of the optical potential as
\begin{equation}
   \sigma_{\rm abs} = \int d^2 b\,\left(1 - e^{\frac{2}{v}\int\limits_{-\infty}^{+\infty} dz\,{\rm Im}\,U(\bm{b},z)}\right)~.      \label{sigma_abs_U}
\end{equation}

Keeping only the contribution of the central interaction, Eq.(\ref{A_tau}), the phase-shift
function is expressed as follows:
\begin{equation}
   \chi_N(b) = \frac{i}{2} \sum_\tau  \sigma_\tau (1-i\alpha_\tau) \int \frac{d^2q_{t}}{(2\pi)^2}
                    e^{i\bm{q}_{t} \cdot \bm{b}}  S_\tau(-\bm{q}_{t}) e^{-\beta_\tau q_{t}^2/2}~.  \label{chi_N_central}
\end{equation}
If we now approximately set $\beta_\tau=0$, which is a reasonable assumption if the form factors of the nucleon densities
drop quickly with momentum transfer, then we have
\begin{equation}
     \chi_N(b) = \frac{i}{2} \sum_\tau  \sigma_\tau (1-i\alpha_\tau) T_\tau(b)~,   \label{chi_N_beta0}
\end{equation}
which leads to the well known semiclassical formula for the absorption cross section:
\begin{equation}
   \sigma_{\rm abs} = \int d^2 b\,
   \left(1 - e^{-\sum\limits_{\tau=n,p} \sigma_\tau T_\tau(b)}\right)~.    \label{sigma_abs_class}
\end{equation}

\section{Numerical results}
\label{NumRes}

For concrete calculations we need to fix the parameters of the elastic $\bar p N$ and $pN$ amplitudes
which are expressed by Eqs.(\ref{f_tau})-(\ref{C}). 
Thus, before discussing numerical results we describe the default settings which are used unless
the modification is explicitly specified. 

\begin{figure}
\begin{center}
   \includegraphics[scale = 0.60]{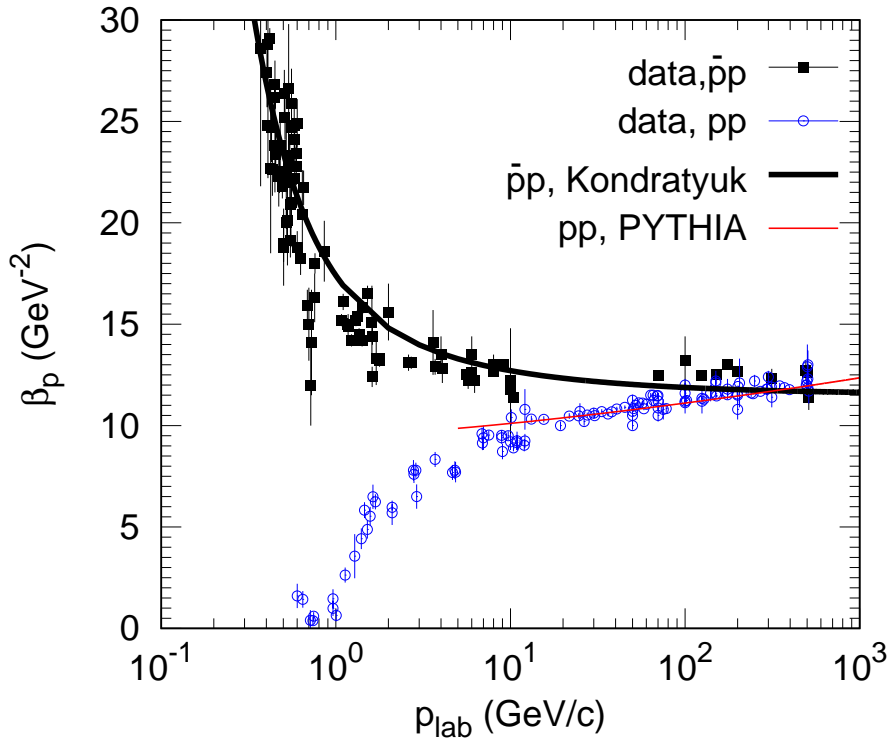}
\end{center}
\caption{\label{fig:slopes} The slopes of transverse momentum transfer dependence of the elastic $\bar pp$ and $pp$ amplitudes
  vs beam momentum. Thick solid line corresponds to the $\bar pp$ amplitude as parameterized by Eq.(\ref{beta_pbarp}).
  Thin solid line displays the $pp$ slope in the PYTHIA parameterization, Eq.(\ref{beta_pp}).
  The data points are taken from the compilation in ref. \cite{Okorokov:2015bha} (low momentum transfer data sets).}
\end{figure}

{\bf $\bm{\bar p N}$ amplitude.}
The total $\bar p p$ cross section is parameterized as follows:
\begin{equation}
\sigma_p(p_{\rm lab}) =
     \left\{
     \begin{array}{ll}
        \exp\{4.5485 \exp[-0.0601 \ln(T_{\rm lab})]\}  & \mbox{for}~p_{\rm lab} < 5.92 \\
         38.4+77.6p_{\rm lab}^{-0.64}+0.26\ln^2(p_{\rm lab})-1.2\ln(p_{\rm lab}) & \mbox{for}~ p_{\rm lab} \geq 5.92  
     \end{array}    
     \right.                         \label{sig_pbarp^tot}
\end{equation} 
with $p_{\rm lab}$ in GeV/c and the cross section in mb. $T_{\rm lab}=\sqrt{p_{\rm lab}^2+m_p^2}-m_p$ is the antiproton kinetic energy in GeV.
Here, at lower and high beam momenta, respectively, the parameterizations from ref. \cite{Clover:1982qq} 
and from PDG of ref. \cite{PhysRevD.50.1173} are used.

The ratio of the real-to-imaginary part of the $\bar p p$ forward scattering amplitudes, $\alpha_p$, is set as
\begin{equation}
   \alpha_p=
    \left\{
    \begin{array}{ll}
       0.2 &  \mbox{for}~p_{\rm lab} < 0.7 \\
       0.1 &  \mbox{for}~0.7 \leq  p_{\rm lab} < 0.9 \\
       0.06&  \mbox{for}~0.9 \leq  p_{\rm lab} < 1.5
    \end{array}    
    \right.                         \label{alpha_p}
\end{equation}
with $p_{\rm lab}$ in GeV/c. These settings correspond to the PDG systematics \cite{Agashe:2014kda} 
at low beam momenta. At higher beam momenta, we adopt for $\alpha_p$ the Regge-Gribov fit at $\sqrt{s} \geq 5$ GeV 
from PDG \cite{Agashe:2014kda}.

The slope of the momentum dependence is obtained from the parameterization of ref. \cite{Kondratyuk:1986cq}:
\begin{equation}
    \beta_p = (0.67+0.35/k_{\bar p p})^2~,             \label{beta_pbarp}
\end{equation} 
where $k_{\bar p p}=\sqrt{m_pT_{\rm lab}/2}$ is the $\bar p p$ c.m. momentum in units of fm$^{-1}$
and $\beta_p$ in fm$^2$. As we see from Fig.~\ref{fig:slopes}, Eq.(\ref{beta_pbarp}) provides a good description
of the empirical $\bar pp$ slope in a broad beam momentum region from $\sim 500$ MeV/c up to $10^3$ GeV/c.
At low beam momenta, $p_{\rm lab} \ltsim 1$ GeV/c, the $\bar p p$ elastic amplitude is sharply peaked
at forward c.m. scattering angles, in contrast to almost isotropic $pp$ amplitude. 
This is because in the $\bar p p$ case the contribution of lower partial waves (small impact parameters) is suppressed
by annihilation. With increasing beam momentum above $\sim10^2$ GeV/c the slopes of $\bar p p$ and $pp$ elastic amplitudes
become close to each other \footnote{According to the Regge theory \cite{Collins}, at even higher beam momenta both $\bar p p$ and $pp$
slopes increase as $\beta_p= \mbox{const} + 2\alpha_p^\prime\log(s/s_0)$, where $\alpha_p^\prime=0.2$ is the slope of the Pomeron
trajectory and $s_0 \simeq 1$ GeV$^2$. This is confirmed by the recent TOTEM data \cite{Antchev:2013gaa} which give $\beta_p=19.9\pm0.3$ GeV$^{-2}$
for $pp$ collisions at $\sqrt{s}=7$ TeV.}.

To minimize the number of free parameters, we always set the neutron parameters equal to the proton ones, i.e.
$\sigma_n=\sigma_p$, $\alpha_n=\alpha_p$ and $\beta_n=\beta_p$.

For the spin-orbit $\bar p N$ interaction we use the following settings. 
At $p_{\rm lab}=0.608$ GeV/c, we set the strength $\sigma_c=1.40~\mbox{fm}^3$.
At higher beam momenta, $\sigma_c$ is obtained by rescaling its value at 0.608 GeV/c:
\begin{equation}
  \sigma_c(p_{\rm lab}) =\sigma_c(0.608~\mbox{GeV/c}) \frac{\sigma_p(p_{\rm lab})}{\sigma_p(0.608~\mbox{GeV/c})}~.    \label{sigma_c_plab}
\end{equation}
For simplicity, we assume that $\beta_c=\beta_p$ and $\alpha_c=0$. These settings, in general, agree with refs. \cite{Martin:1988wv,Tan:1989gd} 
and, as we will see below, well confirmed by the comparison to data.

{\bf $\bm{pN}$ amplitude.} The total $pp$ and $pn$ cross sections, $\sigma_p$ and $\sigma_n$, and
the ratio of the real-to-imaginary part, $\alpha_p$, are taken from the Regge-Gribov fit at $\sqrt{s} \geq 5$ GeV from PDG \cite{Agashe:2014kda}.  
The slope parameter of the $pp$ amplitude is taken from the PYTHIA model as described in ref. \cite{Falter:2004uc}:
\begin{equation}
  \beta_p = 5.0 + 4s^{0.0808}~,             \label{beta_pp}
\end{equation}
where $s=2m_p(\sqrt{p_{\rm lab}^2+m_p^2}+m_p)$ is the square of the $\bar p p$ c.m. energy in units of GeV$^2$
and $\beta_p$ in GeV$^{-2}$. From Fig.~\ref{fig:slopes} we see that the parameterization (\ref{beta_pp}) well describes the empirical $pp$ slope
above $p_{\rm lab} \sim 9$ GeV/c. It is assumed in calculations that $\alpha_n=\alpha_p$ and $\beta_n=\beta_p$.
The spin-orbit $pN$ interaction is neglected.

\subsection{Elastic and quasielastic scattering, polarization}
\label{results_El}

\begin{figure}
\begin{center}
   \includegraphics[scale = 0.40]{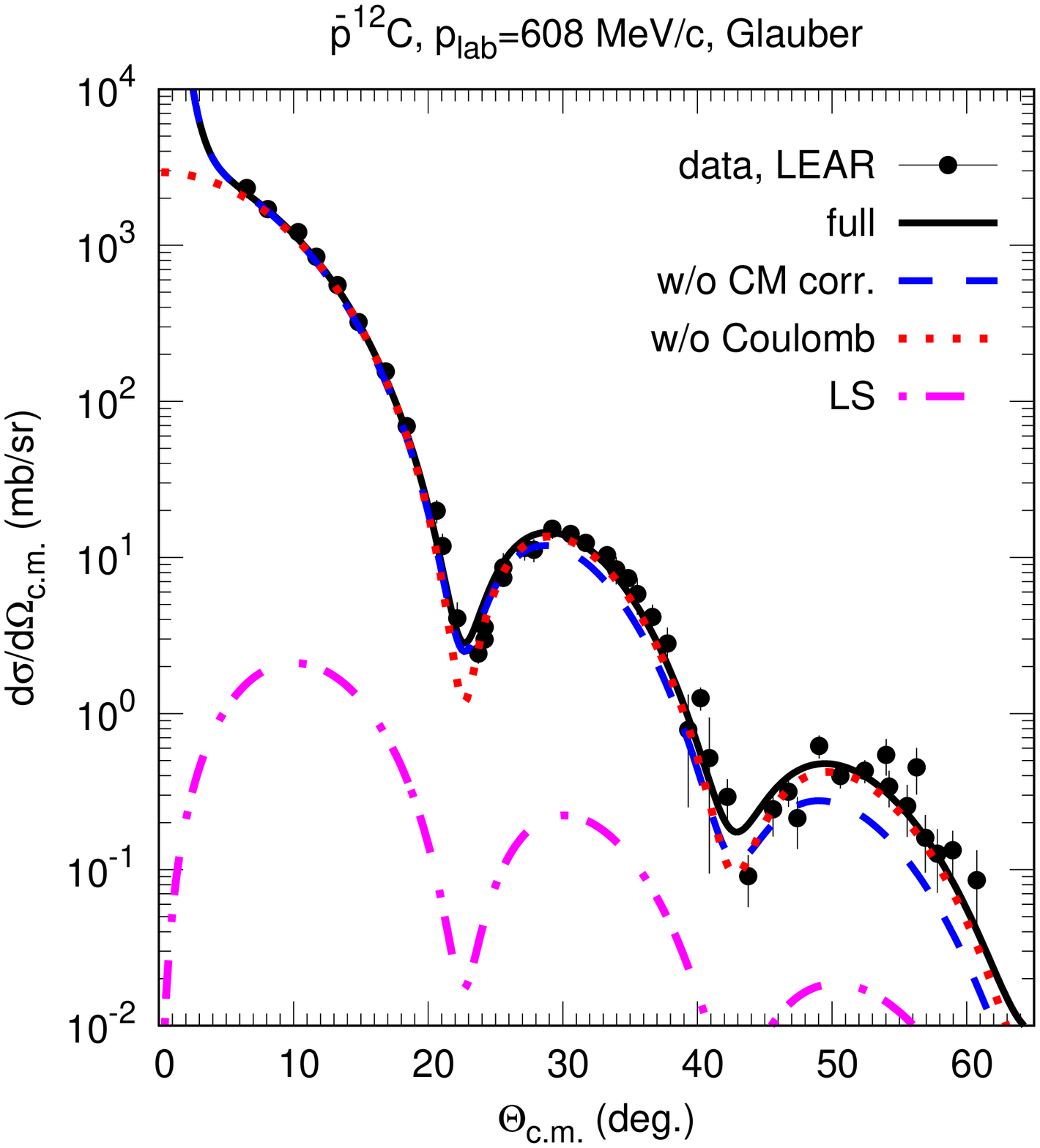}
   \includegraphics[scale = 0.40]{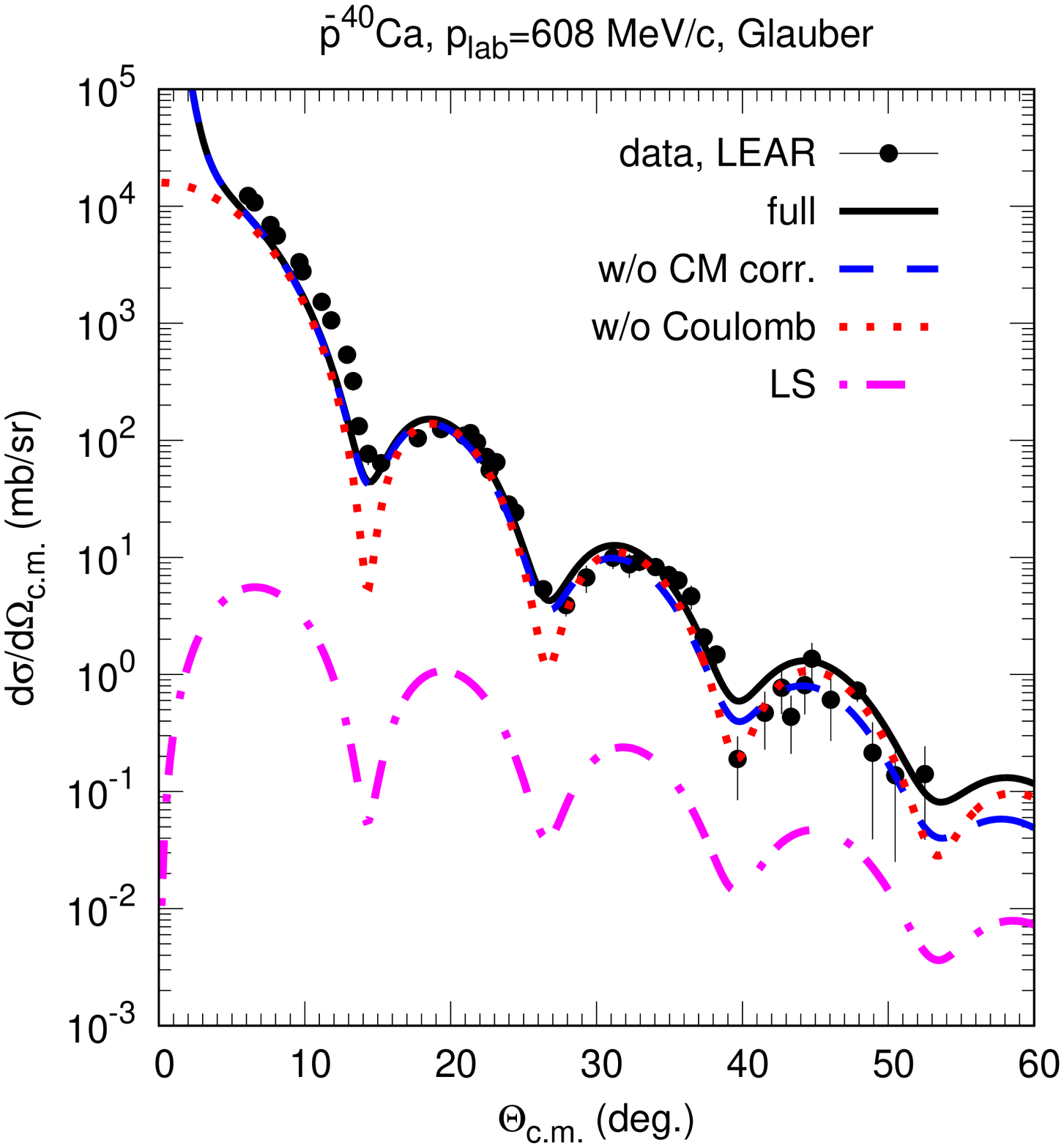}
   \includegraphics[scale = 0.40]{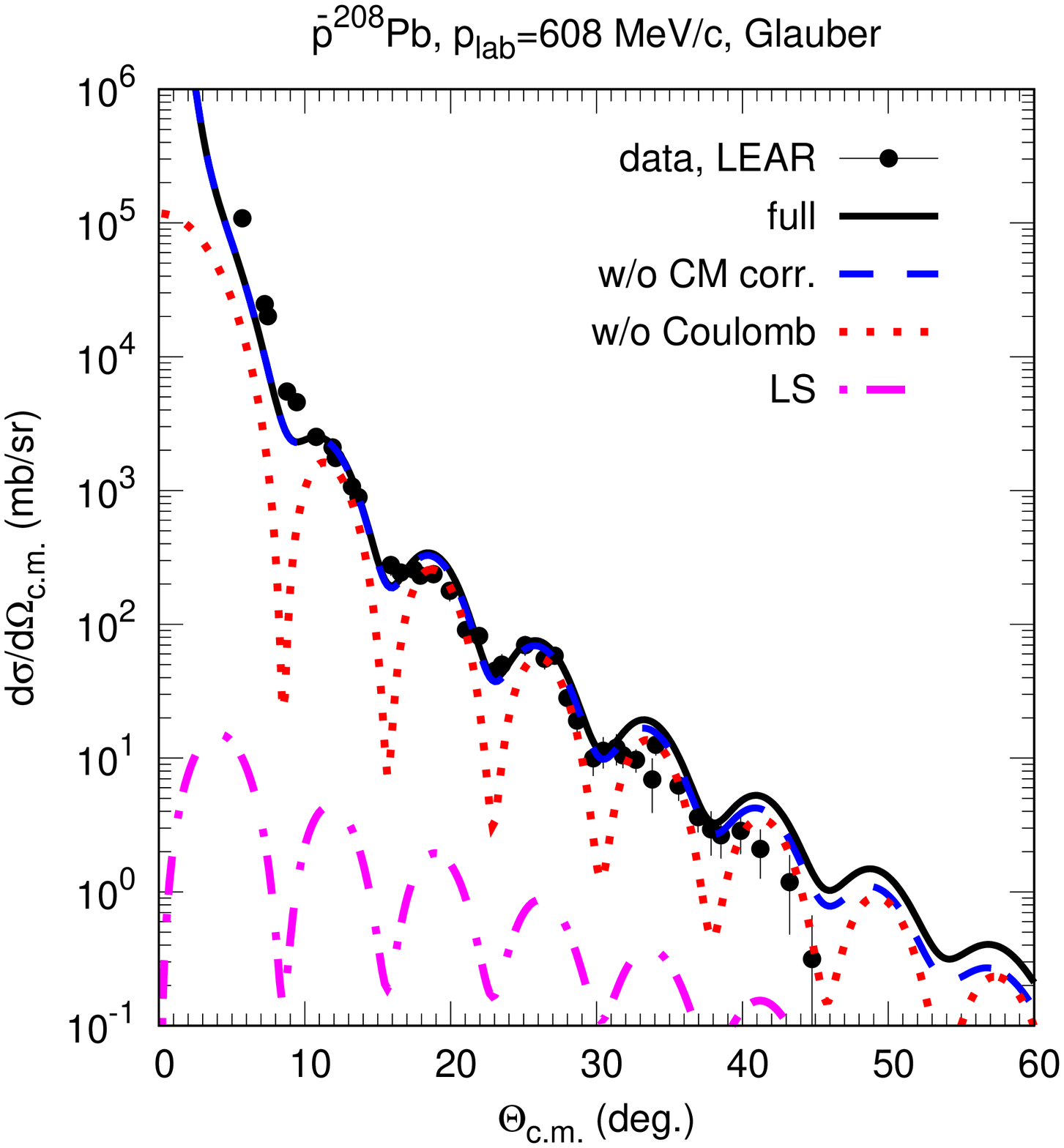}
\end{center}
\caption{\label{fig:dsigdOmega_LEAR} Angular differential cross section of $\bar p$ elastic scattering at 608 MeV/c on $^{12}$C,
  $^{40}$Ca, and $^{208}$Pb. Full GM calculation is shown by solid line. The dashed and dotted lines show, respectively,
  the results without recoil correction ($H_{\rm cm}(q)=1$, Eq.(\ref{H_cm}))
  and without Coulomb correction ($\xi=0$, Eq.(\ref{xi})). The dot-dashed line shows the contribution of the spin-orbit
  amplitude $G$ to the differential cross section, Eq.(\ref{dsigmadOmega_cm}). Experimental data are from ref. \cite{Garreta:1984rs}.}
\end{figure}
In Fig.~\ref{fig:dsigdOmega_LEAR} we show the comparison of our calculations with LEAR data \cite{Garreta:1984rs} for
the angular differential cross sections of antiproton elastic scattering at $T_{\rm lab}=180$ MeV ($p_{\rm lab}=608$ MeV/c) 
on carbon, calcium and lead targets.
Apart from the full GM calculation, also the effects of the various model
ingredients are shown. As we see, the Coulomb scattering  entirely dominates at small angles and is also important at diffractive
minima for heavy nuclei. The c.m. (recoil) correction is important for light nuclei at large angles:
it reaches $\sim 75\%$ for $^{12}$C at 50$^0$.
In agreement with ref. \cite{Tan:1989gd}, we also see that the relative importance of the spin-orbit interaction grows with
scattering angle, although it always provides orders of magnitude smaller contribution to the differential elastic scattering
cross section as compared to the central interaction.
The full GM calculation is in a good agreement with experimental data. It is also in line with the earlier
GM calculations of refs. \cite{Dalkarov:1986tu,Tan:1989gd} for the same reactions and with similar model parameters.

\begin{figure}
\begin{center}
   \includegraphics[scale = 0.40]{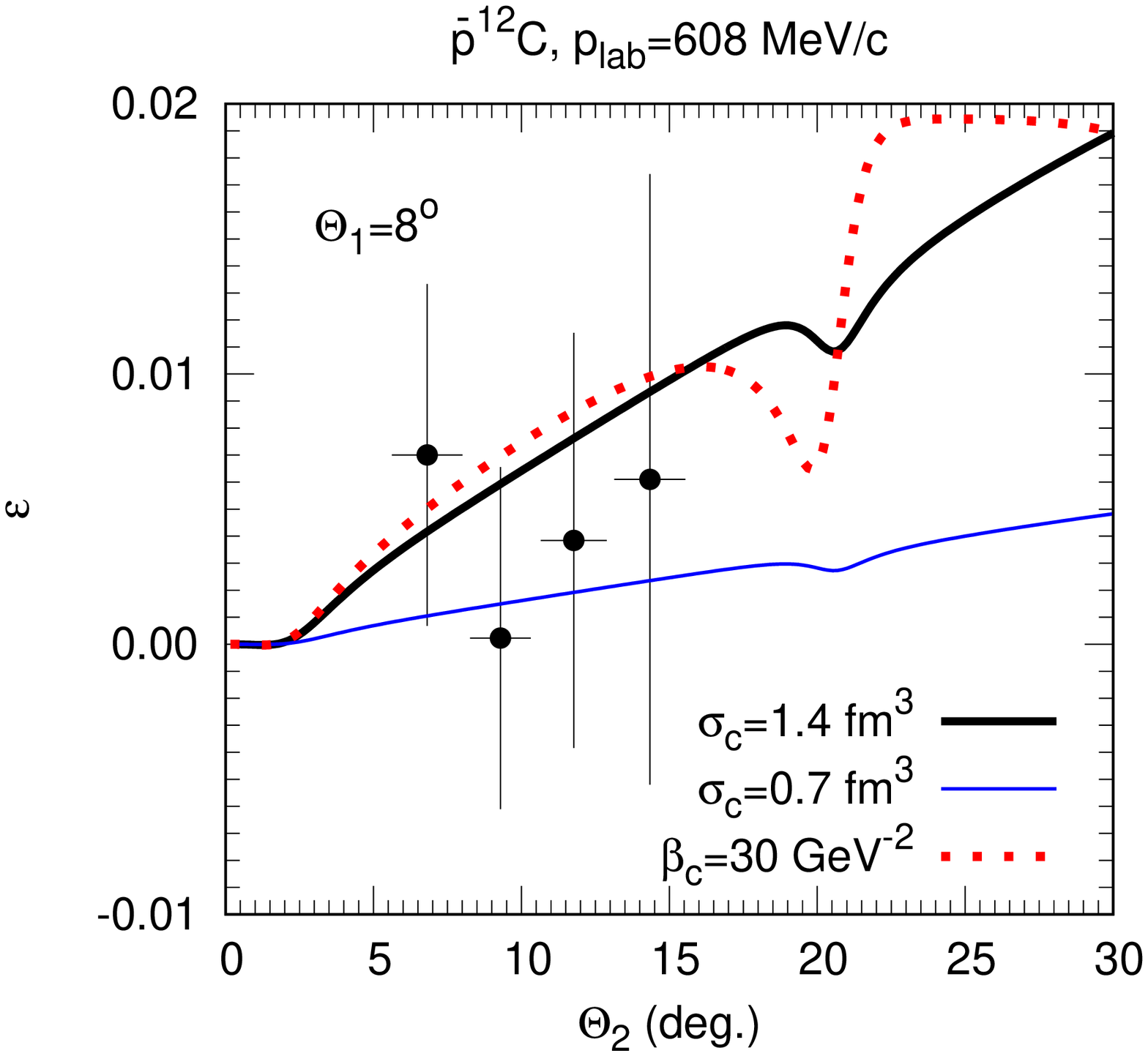}
   \includegraphics[scale = 0.40]{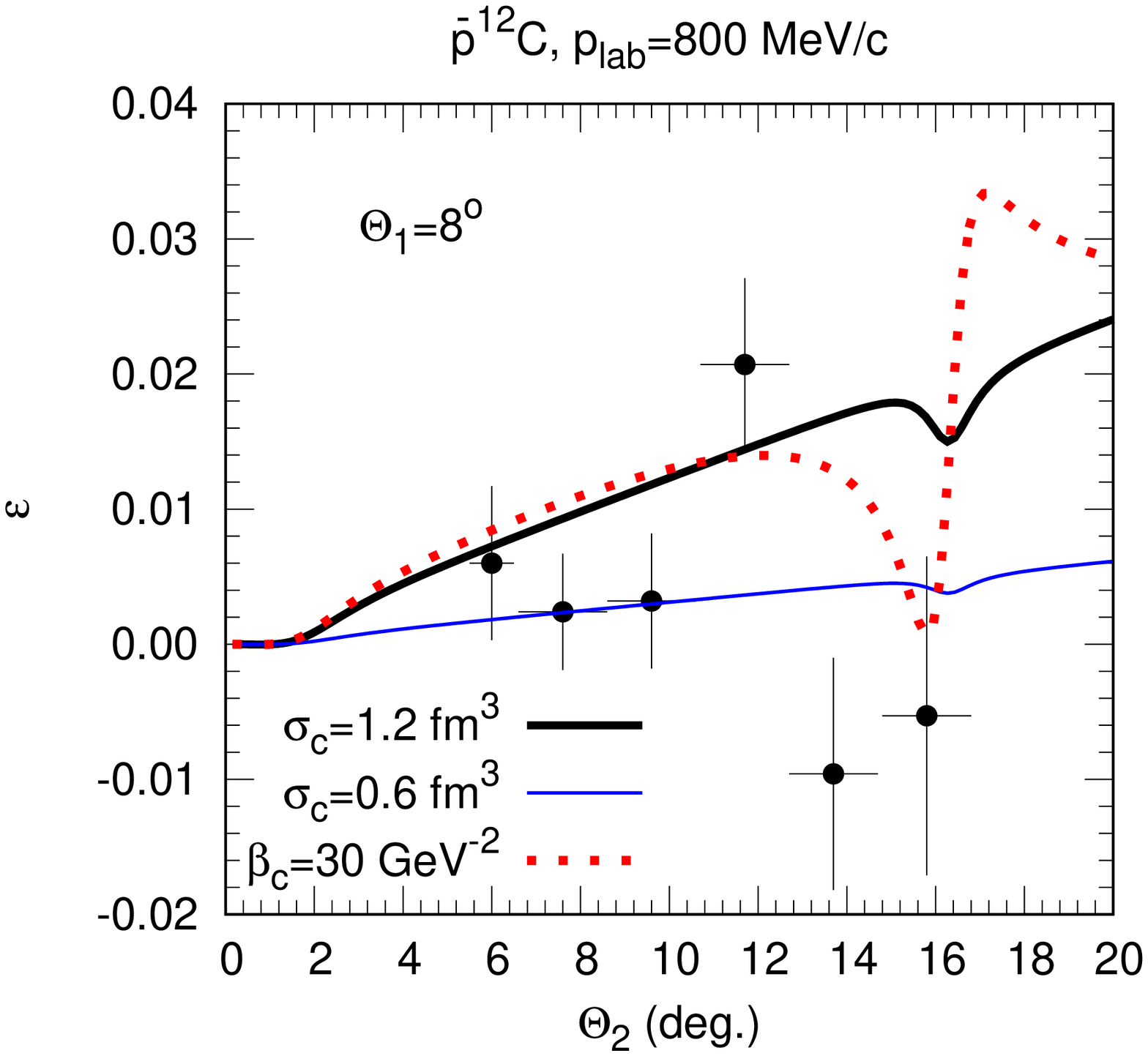}
   \includegraphics[scale = 0.40]{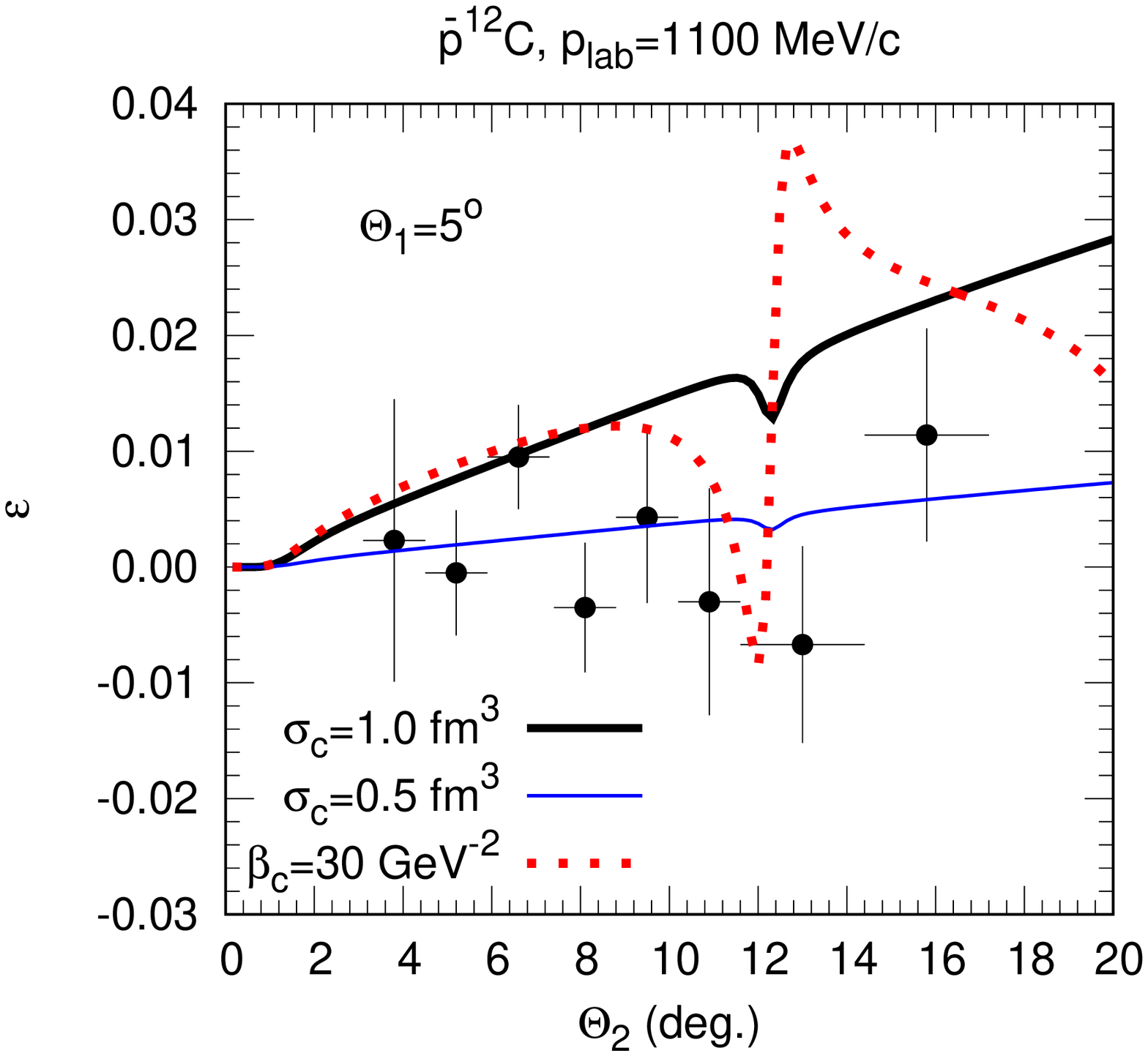}
   \includegraphics[scale = 0.40]{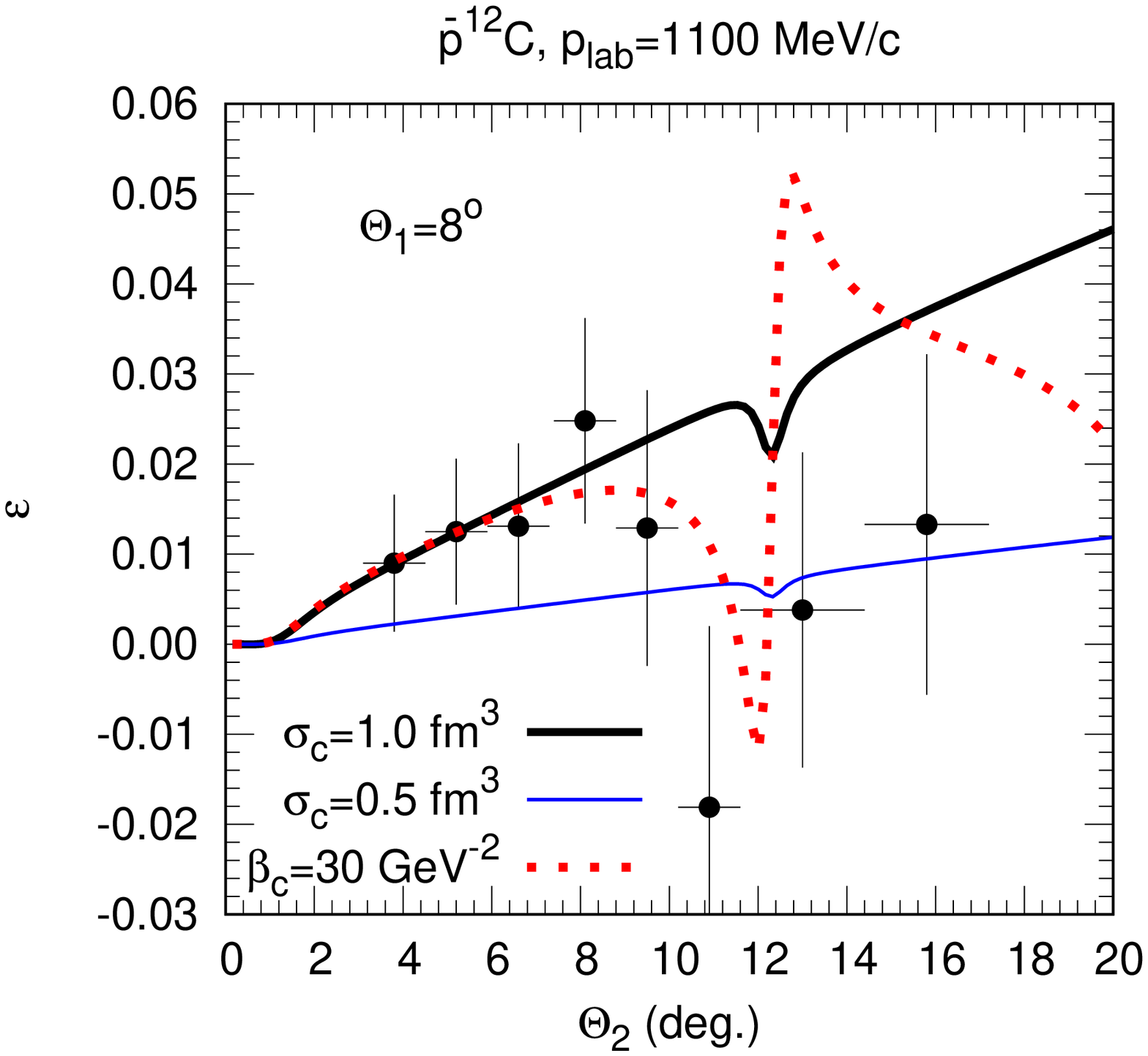} 
\end{center}
\caption{\label{fig:eps_LEAR} The azimuthal asymmetry, Eq.(\ref{asymmetry}), at fixed first
  polar laboratory scattering angle, $\Theta_1$, as indicated, as a function of the second polar laboratory scattering angle, $\Theta_2$,
  for the antiproton double elastic scattering on two carbon targets at $p_{\rm lab}=608,~800$ and $1100$ MeV/c.
  Thick and thin solid lines are computed with the strength of a spin-orbit interaction $\sigma_c(608~\mbox{MeV/c})=1.4$ and $0.7~\mbox{fm}^3$,
  respectively. Dotted lines are computed with the larger slope of the spin-orbit interaction, $\beta_c=30$ GeV$^{-2}$. 
  Experimental data are taken from ref. \cite{Martin:1988wv}.}
\end{figure}
Shown in Fig.~\ref{fig:eps_LEAR} is the azimuthal asymmetry, which is defined as a product of the two polarizations,
\begin{equation}
  \varepsilon(\Theta_1,\Theta_2) = P(\Theta_1) P(\Theta_2)~.    \label{asymmetry}
\end{equation}
Experimentally (see ref. \cite{Martin:1988wv}), this quantity has been extracted from the distribution of the double-scattering events
over azimuthal angle $\phi$ between the two scattering planes,
\begin{equation}
  \frac{dN}{d\phi} \propto 1 +  \varepsilon(\Theta_1,\Theta_2) \cos\phi~.    \label{dNdphi}
\end{equation}
The LEAR data \cite{Martin:1988wv} seems to indicate the positive polarization signal growing with
the second polar scattering angle.
However, due to large errorbars the data can be described with the strength of the spin-orbit interaction in a quite
wide range. If one takes the negative values of $\varepsilon$ seriously -- they
can be better fit by calculation with a larger slope, $\beta_c=30$ GeV$^{-2}$, of the spin-orbit interaction
as compared to the default values $\beta_c=21.2,~18.8$ and $16.9$ GeV$^{-2}$ for $p_{\rm lab}=608,~800$ and $1100$ MeV/c, respectively.
The value $\beta_c=30$ GeV$^{-2}$ is close to the one from the Dover-Richard potential model (see ref. \cite{Dover:1980pd}
and Appendix of ref. \cite{Martin:1988wv}).

A more refined statistical analysis of ref. \cite{Martin:1988wv}
taking into account the number of accumulated events for each data point and combining the measurements 
at $p_{\rm lab}=608,~800$ and $1100$ MeV/c gives the polarization in $\bar p C$ scattering increasing linearly
with momentum transfer:
\begin{equation}
   P=a_c q~,        \label{P_Martin}
\end{equation}
where $a_c=+0.72$\protect\raisebox{-0.5ex}{$\stackrel{\scriptstyle +0.09}{\scriptstyle -0.10}$} $(\mbox{GeV/c})^{-1}$
is an energy independent coefficient.
\begin{figure}
\begin{center}
   \includegraphics[scale = 0.60]{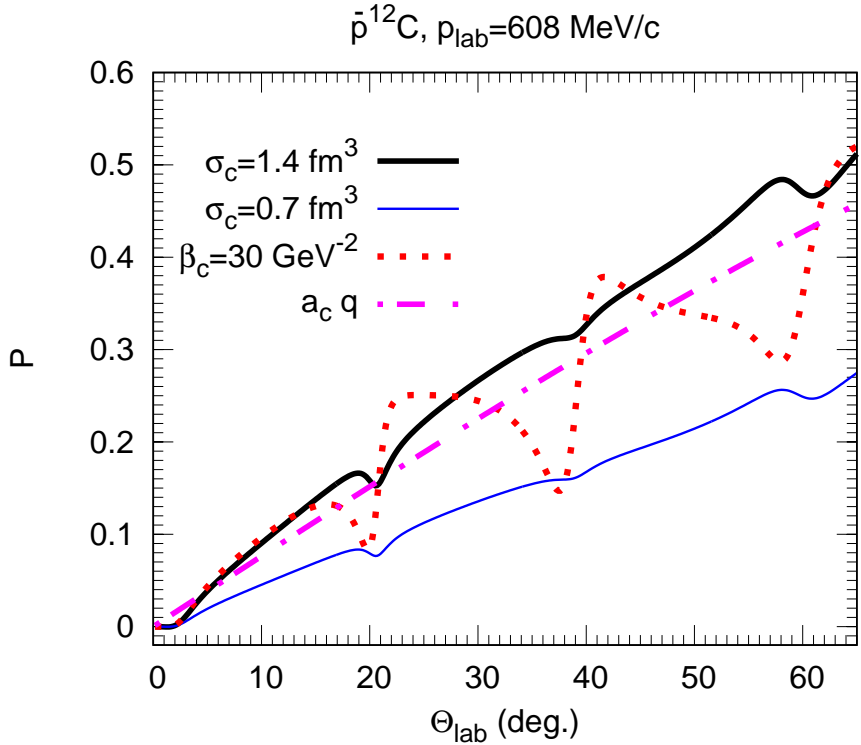}
\end{center}
\caption{\label{fig:pol_LEAR} Polarization as a function of the c.m. scattering angle for $\bar p$ elastic scattering on $^{12}$C
  at $p_{\rm lab}=608$ MeV/c. The thick and thin solid lines show the GM calculations with the different values
  of $\sigma_c$, as indicated. The dotted line displays the result with the larger slope of the spin-orbit interaction, $\beta_c=30$ GeV$^{-2}$.
The dot-dashed line shows the fit, Eq.(\ref{P_Martin}), of experimental data from ref. \cite{Martin:1988wv}.}
\end{figure}
As visible from Fig.~\ref{fig:pol_LEAR}, Eq.(\ref{P_Martin}) is close to the calculation using the upper boundary for the spin-orbit interaction
strength.
However, the calculation with stiffer slope of the spin-orbit interaction produces more pronounced structure of the polarization pattern
and it is hard to conclude from Fig.~\ref{fig:eps_LEAR} which parameterization describes the data better.

\begin{figure}
\begin{center}
   \includegraphics[scale = 0.50]{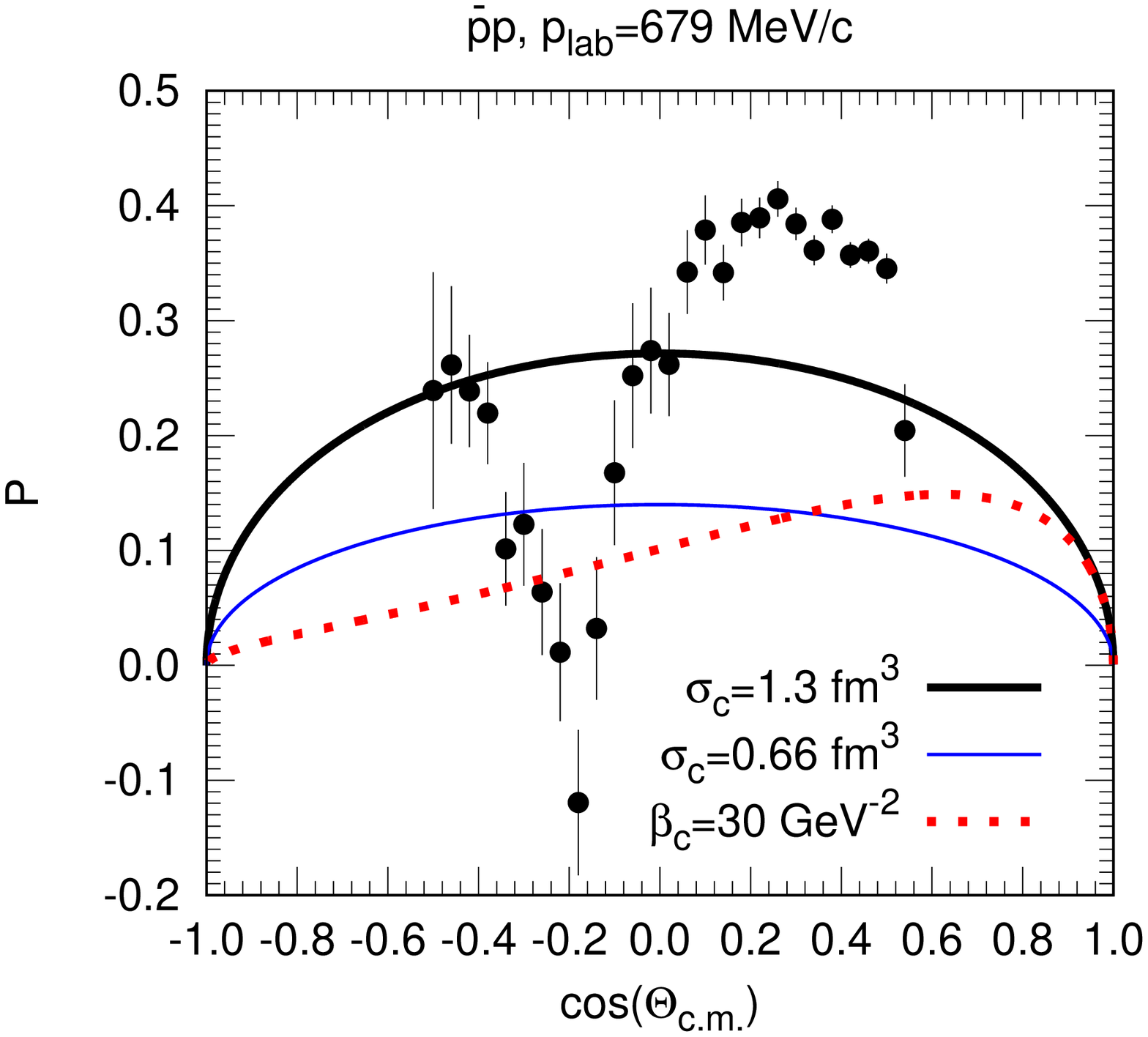}
   \includegraphics[scale = 0.50]{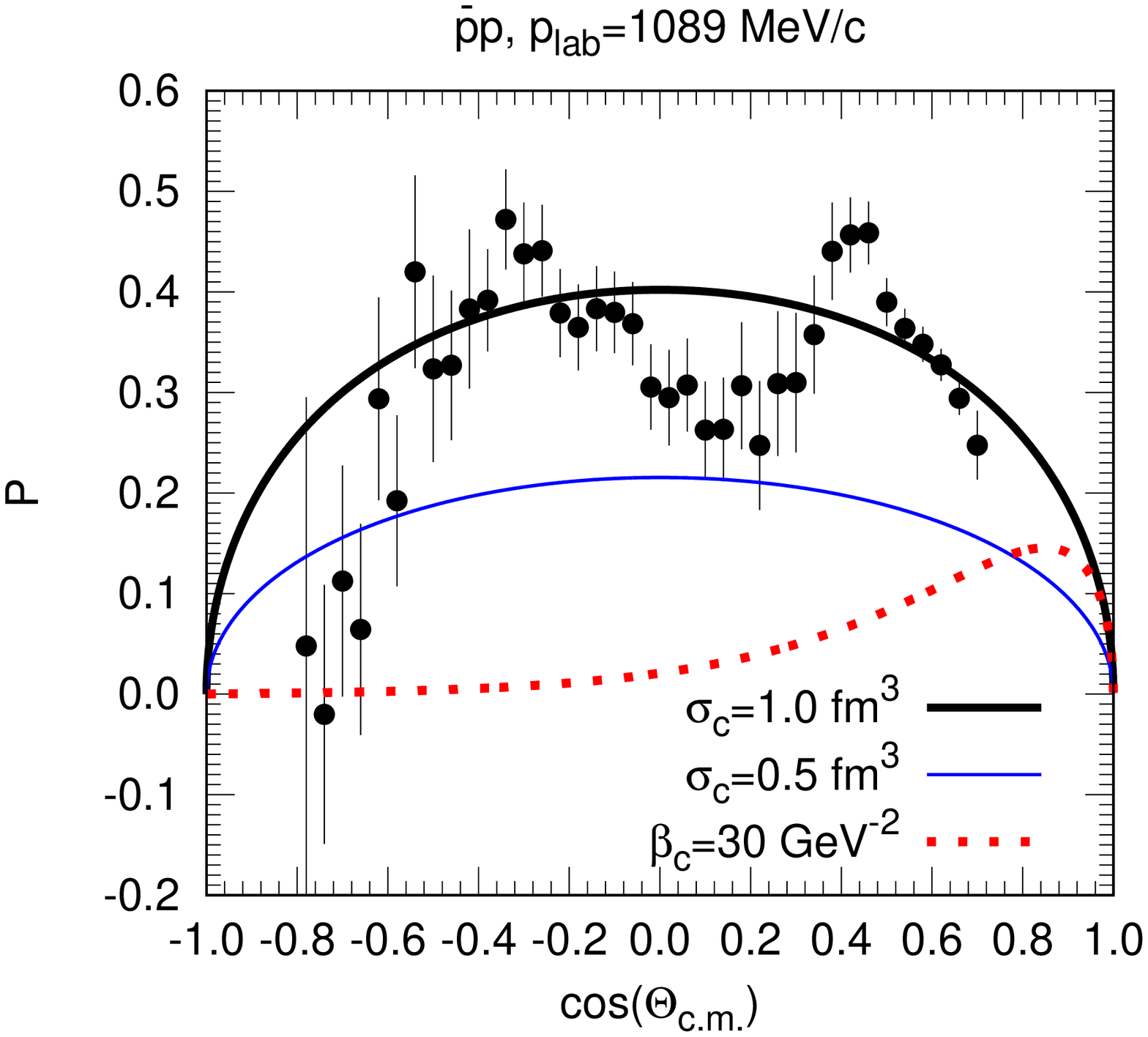}
\end{center}
\caption{\label{fig:pol_pbarp} Polarization as a function of the cosine of the c.m. polar scattering angle for $\bar p p$ elastic scattering
at $p_{\rm lab}=679$ and 1089 MeV/c. Line notations are the same as in Fig.~\ref{fig:eps_LEAR}. 
Experimental data are taken from ref. \cite{Kunne:1988tk}.}
\end{figure}
In such situation, it is natural to go back to the $\bar p p$ elastic scattering.   
Fig.~\ref{fig:pol_pbarp} shows the polarization in the elementary $\bar p p$ collisions calculated according to Eq.(\ref{P_pbarp}).
It is still difficult to judge which set of parameters describes the data at 679 MeV/c better.
An obvious shortcoming of our calculations is that the diffractive structure of $P[\cos(\Theta_{c.m.})]$ is not reproduced.
This is an indication of missing tensor and quadratic spin-orbit contributions to the $\bar p p$ elastic amplitude.    
But, fortunately, these contributions are proportional to the spin of the target nucleon and, thus,
can be disregarded for the spin-zero nuclear targets \cite{Osland:1978vu}.

It is also clear from Fig.~\ref{fig:pol_pbarp} that at 1089 MeV/c the calculation with stiff slope of the spin-orbit interaction
fails to describe the data.
In principle, this discrepancy might be at least partly compensated by tensor and quadratic spin-orbit interactions.
However, having in mind that, according to ref. \cite{Kunne:1988tk}, the Dover-Richard model underpredicts
the analysing power at $p_{\rm lab} \ltsim 1$ GeV/c and contradicts to the data at $p_{\rm lab} > 1$ GeV/c,
it is quite improbable that the calculation with $\beta_c=30$ GeV$^{-2}$ is realistic above 1 GeV/c.
Nevertheless, for completeness, we will include the results of polarization calculations at higher energies
obtained with $\beta_c=30$ GeV$^{-2}$ (see Fig.~\ref{fig:pol_10gevc} below). 
 
\begin{figure}
\begin{center}
   \includegraphics[scale = 0.40]{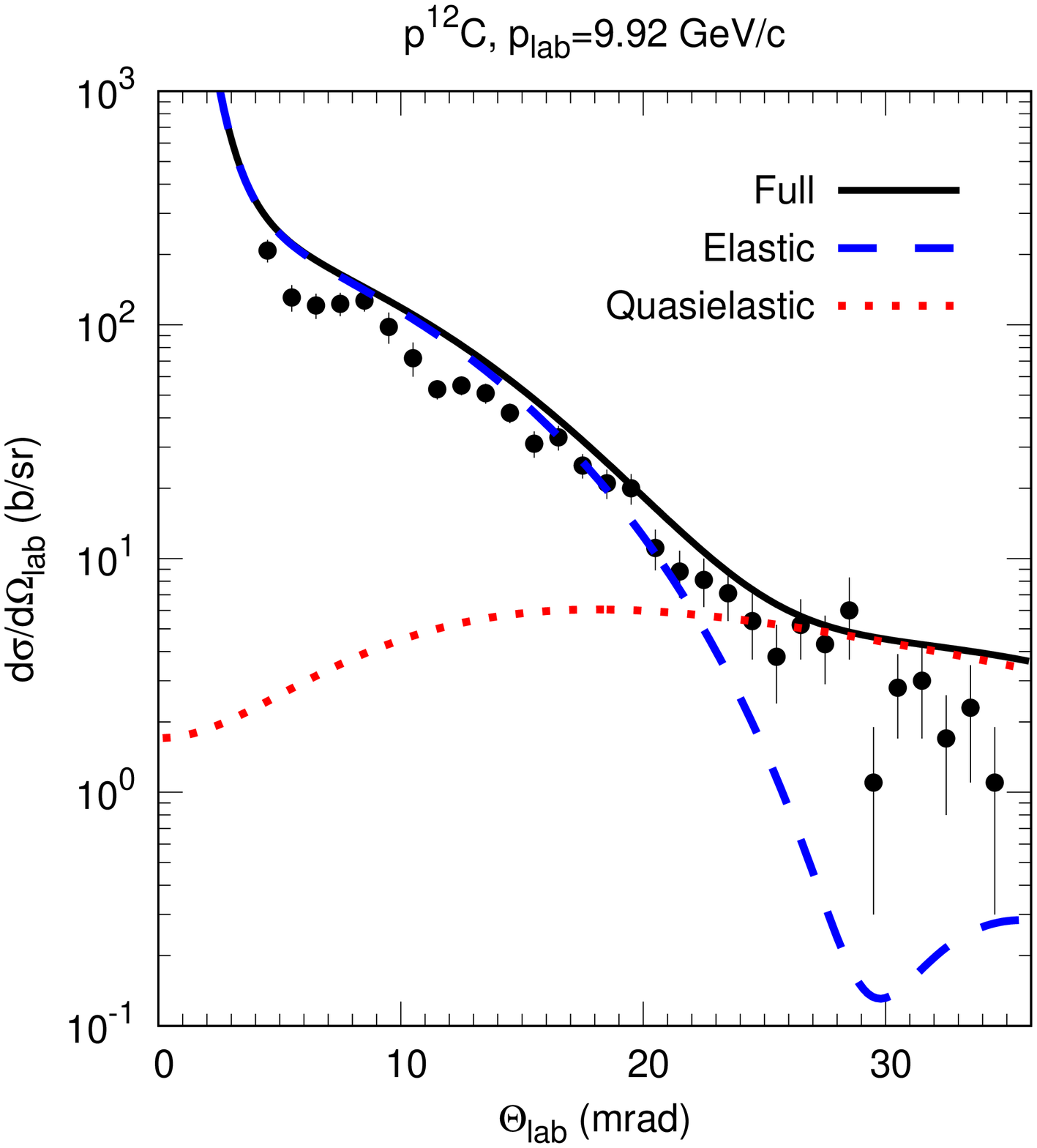}
   \includegraphics[scale = 0.40]{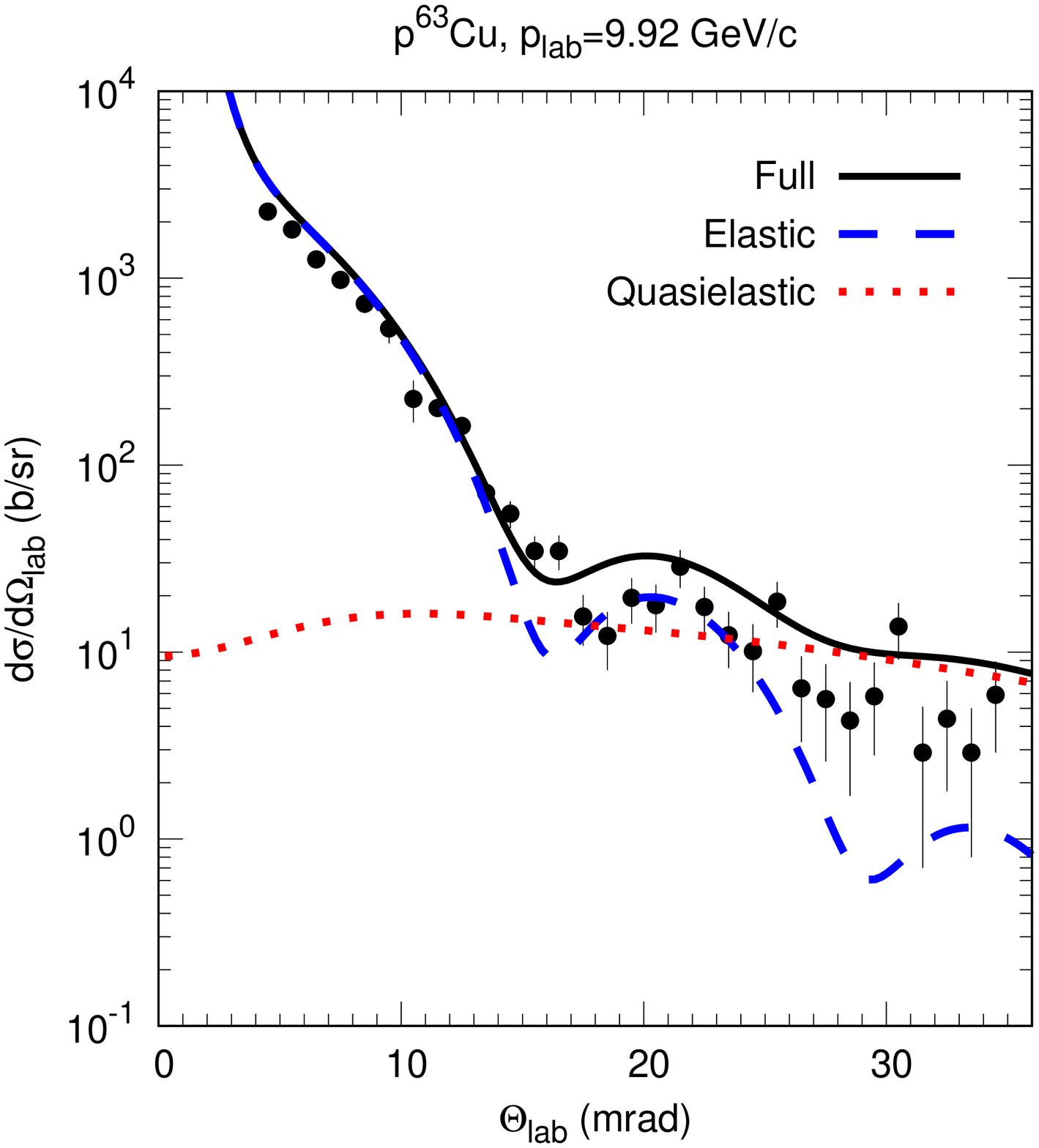}
   \includegraphics[scale = 0.40]{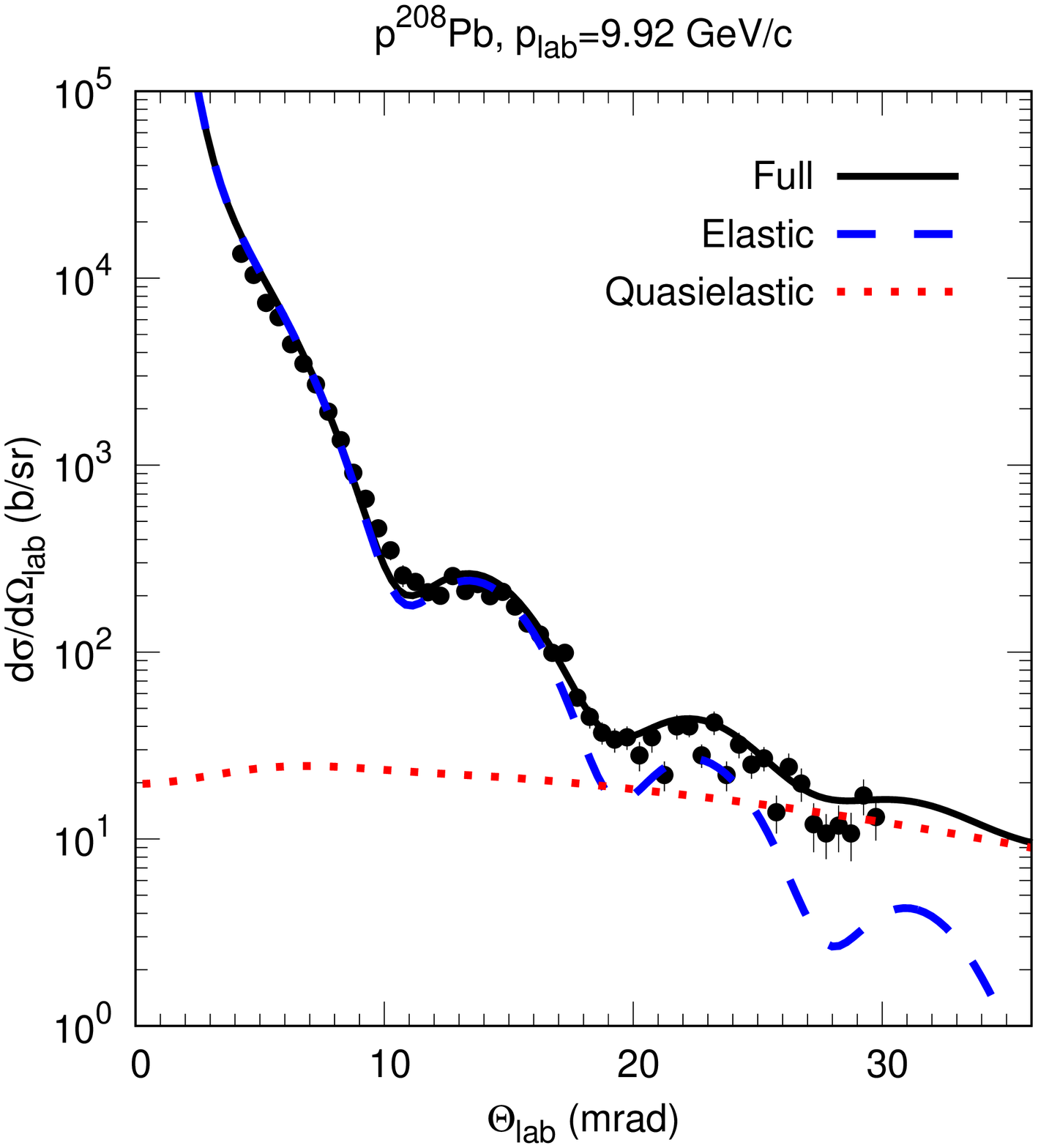}
   \includegraphics[scale = 0.40]{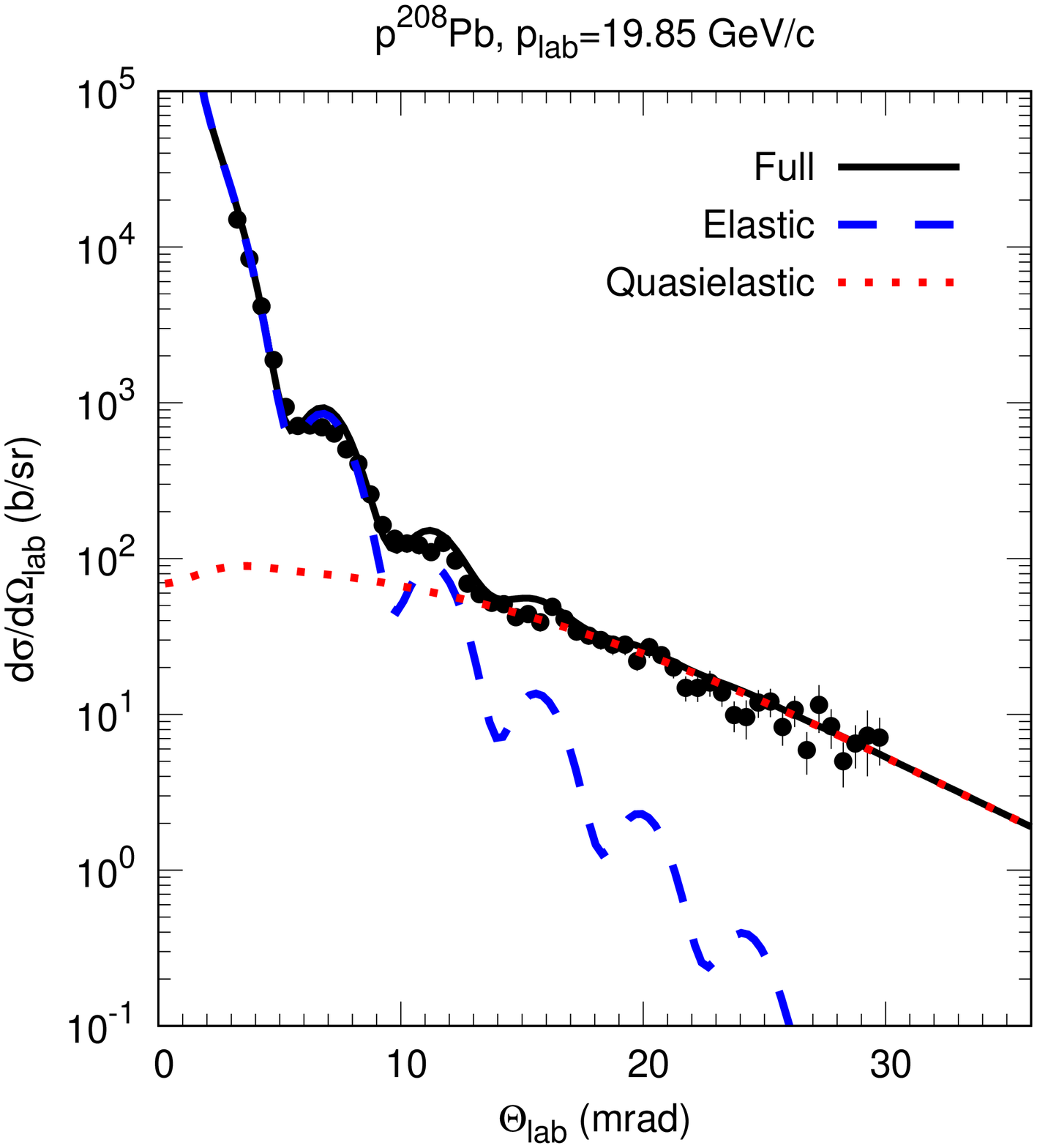}
\end{center}
\caption{\label{fig:AGS} Angular differential cross section of proton scattering on various targets and beam momenta as indicated.
The full GM calculations including elastic and quasielastic contributions are shown by solid lines. Elastic and quasielastic 
contributions are shown by dashed and dotted lines, respectively. The data points are from ref. \cite{Blieden:1974xy}.}
\end{figure}
Let us now turn to the reactions at higher beam momenta where the GM is expected to describe both antiproton- and proton-induced
reactions equally well.
Figure~\ref{fig:AGS} shows the calculated differential cross sections of proton scattering
on carbon, copper and lead at 9.92 GeV/c and on lead at 19.85 GeV/c in comparison to the AGS@BNL data \cite{Blieden:1974xy}.
Since the contribution of quasielastic (QE) scattering is not subtracted in the data \cite{Blieden:1974xy}, 
we included the QE scattering in our calculations too as discussed in subsec. \ref{QE}. As we see, the QE background 
dominates at large scattering angles. For the carbon target, the QE contribution almost totally masks the diffractive structure of elastic 
scattering cross section. Thus, the QE background should be eliminated to see the diffractive minima for light nuclei.

\begin{figure}
\begin{center}
   \includegraphics[scale = 0.50]{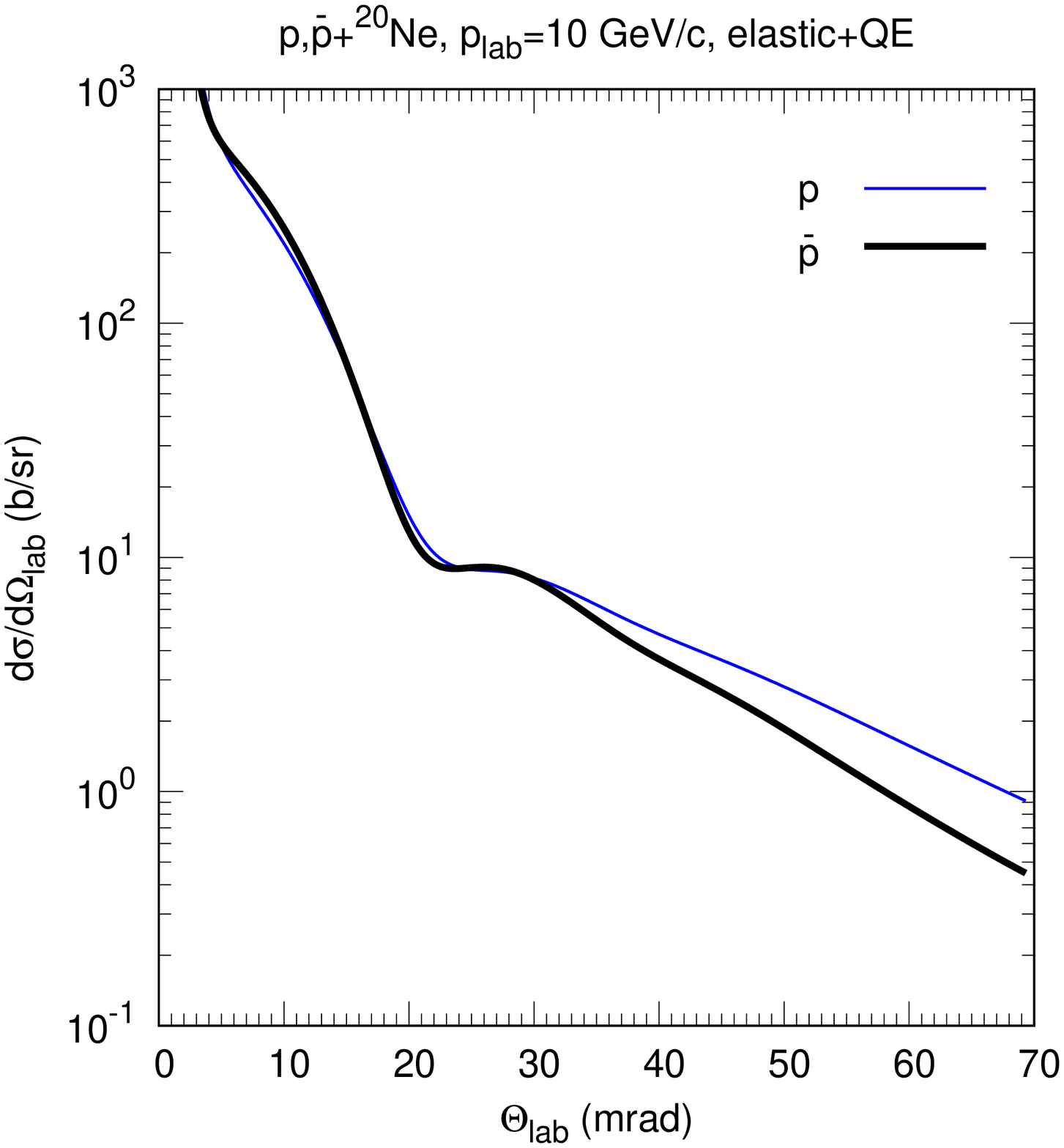}
\end{center}
\caption{\label{fig:p_vs_pbar} Full, i.e. elastic+QE, angular differential cross section of antiproton (thick solid line)
and proton (thin solid line) scattering on $^{20}$Ne at $p_{\rm lab}=10$ GeV/c.}
\end{figure}
\begin{figure}
\begin{center}
   \includegraphics[scale = 0.50]{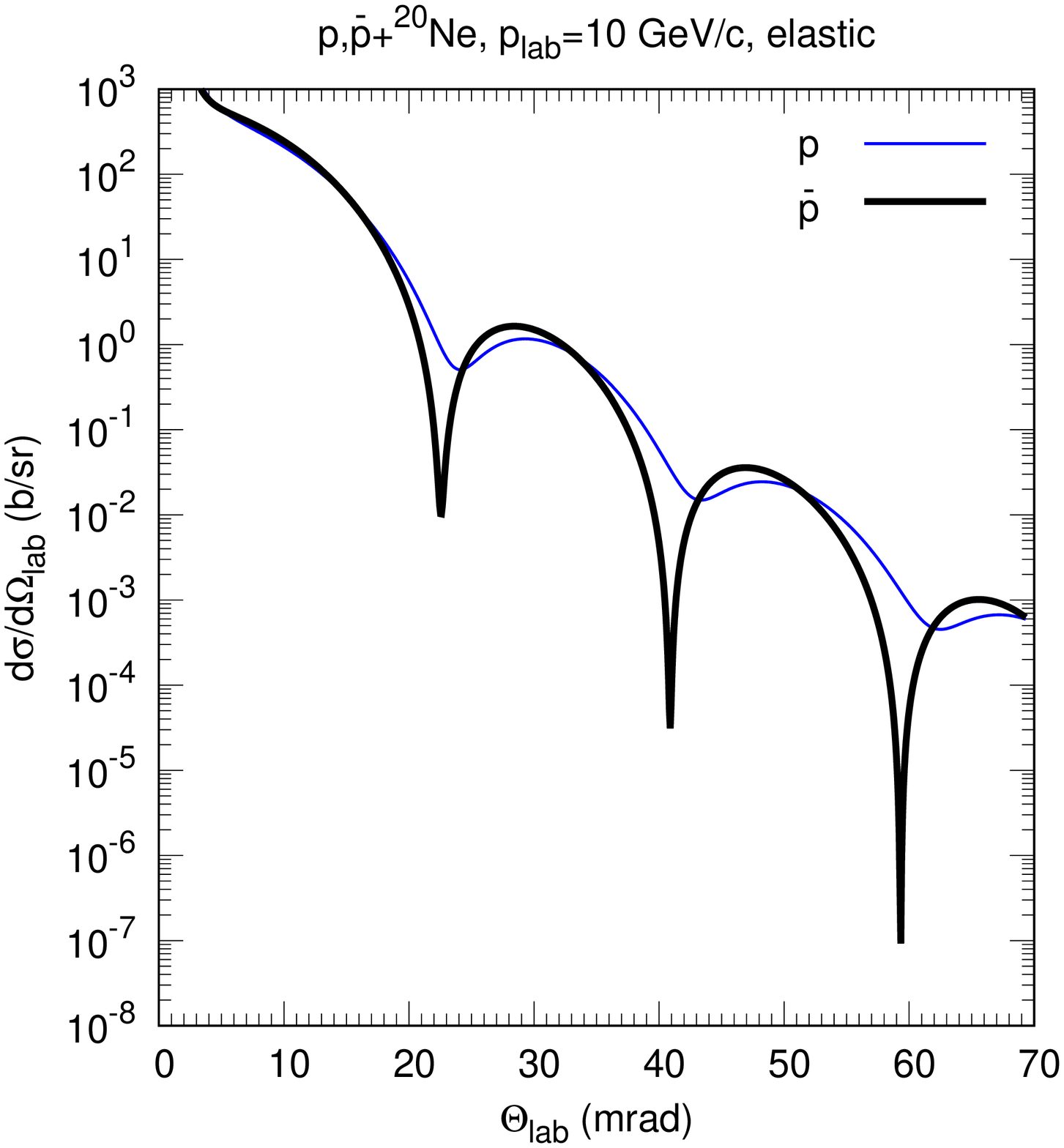}
\end{center}
\caption{\label{fig:p_vs_pbar_EL} Elastic angular differential cross section of antiproton (thick solid line) and proton (thin solid line)
scattering on $^{20}$Ne at $p_{\rm lab}=10$ GeV/c.}
\end{figure}
In Figs.~\ref{fig:p_vs_pbar},\ref{fig:p_vs_pbar_EL} we compare the proton and antiproton angular differential scattering cross sections
at 10 GeV/c on a neon target. For light nuclei, the single-scattering approximation, Eq.(\ref{dsigma_QE_1coll}), can be used to estimate
the QE cross section at large momentum transfers which gives
\begin{equation}
    \frac{d \sigma_{\rm QE}}{d\Omega_{\rm lab}} \propto \mbox{e}^{-\beta_p q^2}~.      \label{sig_QE_est}
\end{equation}
This explains the difference in the slopes of the full cross sections for $p$ and $\bar p$ at large scattering angles
where the QE contribution dominates. 
The most interesting feature, however, is the pronounced difference of elastic cross sections for $p$ and $\bar p$
at the diffractive minima (Fig.~\ref{fig:p_vs_pbar_EL}).
\begin{figure}
\begin{center}
   \includegraphics[scale = 0.50]{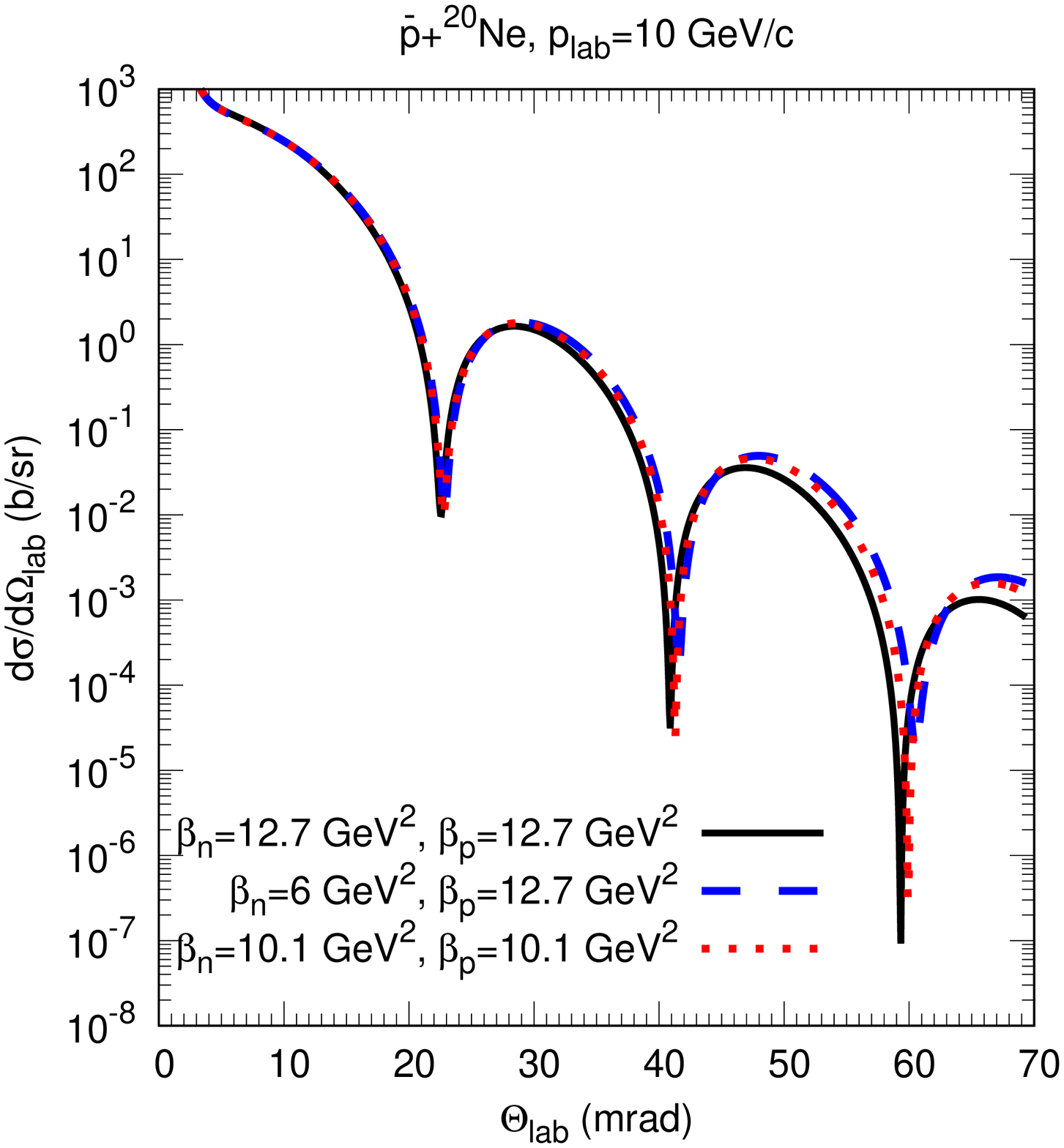}
\end{center}
\caption{\label{fig:beta_dep} Elastic angular differential cross section of antiproton scattering on $^{20}$Ne at $p_{\rm lab}=10$ GeV/c.
The lines show the calculations with different values of the slopes of the momentum dependence, $\beta_n$ and $\beta_p$, as indicated.}
\end{figure}
\begin{figure}
\begin{center}
   \includegraphics[scale = 0.50]{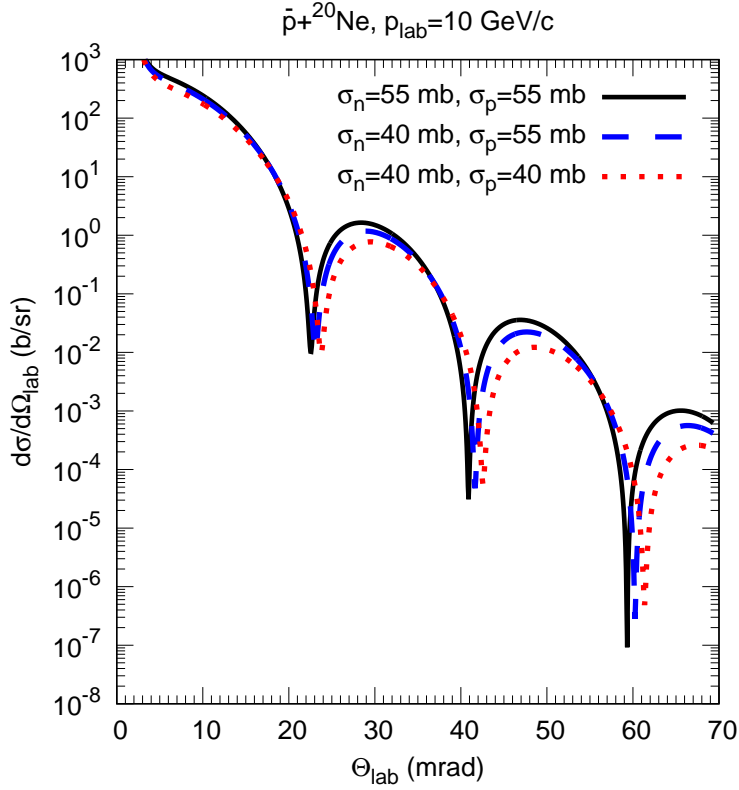}
\end{center}
\caption{\label{fig:sig_dep} Same as in Fig.~\ref{fig:beta_dep} but for the different values of the total cross sections, 
$\sigma_n$ and $\sigma_p$.}
\end{figure}
\begin{figure}
\begin{center}
   \includegraphics[scale = 0.50]{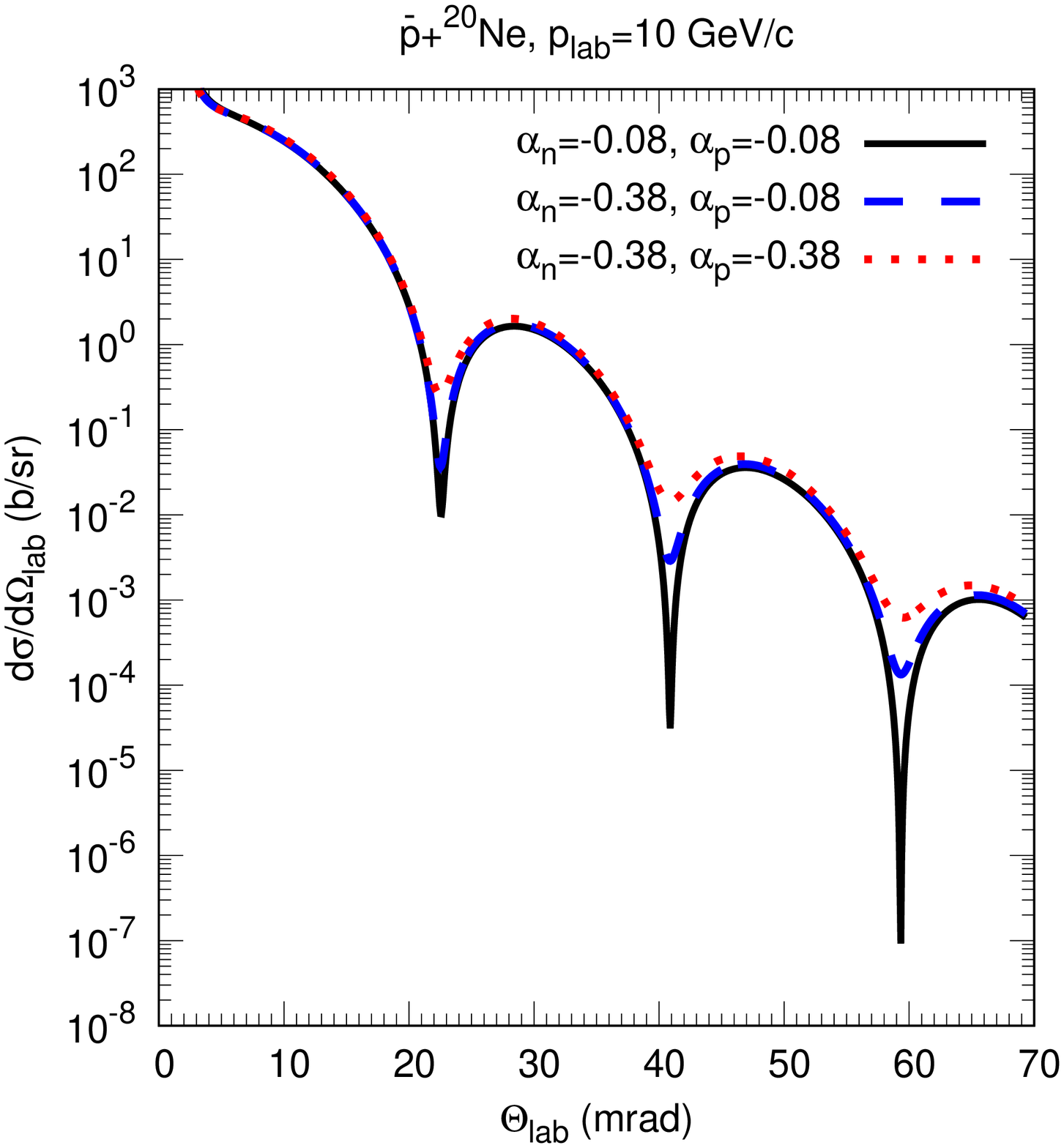}
\end{center}
\caption{\label{fig:alpha_dep} Same as in Fig.~\ref{fig:beta_dep} but for the different values of the ratios of the real-to-imaginary parts, 
$\alpha_n$ and $\alpha_p$.}
\end{figure}
In order to get a deeper insight into this difference, in Figs.~\ref{fig:beta_dep},\ref{fig:sig_dep} and \ref{fig:alpha_dep} we display
the calculations of the differential elastic scattering cross section $\bar p\, ^{20}$Ne at 10 GeV/c with various choices of the slope parameters
of the momentum dependence, total $\bar p N$ cross section, and the ratio of the real-to-imaginary parts of the $\bar p N$ amplitude.
The default values of the parameters of the $\bar pN$ amplitude are $\beta_n=\beta_p=12.7$ GeV$^2$, $\sigma_n=\sigma_p=55$ mb
and $\alpha_n=\alpha_p=-0.08$, while the parameters $\beta_n=\beta_p=10.1$ GeV$^2$, $\sigma_n=\sigma_p=40$ mb and $\alpha_n=\alpha_p=-0.38$
correspond to the $pN$ amplitude at 10 GeV/c.

The influence of the slopes, $\beta_n$ and $\beta_p$, on $d\sigma/d\Omega_{\rm lab}$ is rather weak (Fig.~\ref{fig:beta_dep}).
We observe only a slight shrinking of the diffractive structure and lowering of the diffractive maxima with increasing slopes.
The reason can be understood from Eq.(\ref{chi_N_central}). Larger slopes effectively shrink the nuclear form factors in momentum space and,
therefore, spread the phase-shift function, $\chi_N(b)$, towards larger impact parameters. This acts similar to increasing nuclear radius.

The influence of the total $\bar p N$ cross sections on $d\sigma/d\Omega_{\rm lab}$ is quite significant (Fig.~\ref{fig:sig_dep}).
There is a clean shrinking of the diffractive structure with increasing $\sigma_n$ and $\sigma_p$. This can be again explained
by Eq.(\ref{chi_N_central}), because larger $\sigma$'s extend the profile function of the nucleus, $1-\exp(i\chi_N(b))$,
towards larger impact parameters.

The ratios of the real-to-imaginary parts, $\alpha_n$ and $\alpha_p$, strongly affect the depths of diffractive minima (Fig.~\ref{fig:alpha_dep}).
This is because the finite real part of the $\bar p N$ amplitude produces the oscillating factor, $e^{i\phi}$, with the phase,  
\begin{equation}
   \phi \sim \frac{1}{2}\sum_\tau \sigma_\tau \alpha_\tau T_\tau(b)~,
\end{equation}
in the profile function of the nucleus.
The resulting amplitude on the nucleus is obtained by folding in the space of transverse momentum transfer the amplitude
with $\alpha_\tau=0$ with the Fourier transform of this factor which results in washing-out of the diffractive minima.

\begin{figure}
\begin{center}
   \includegraphics[scale = 0.60]{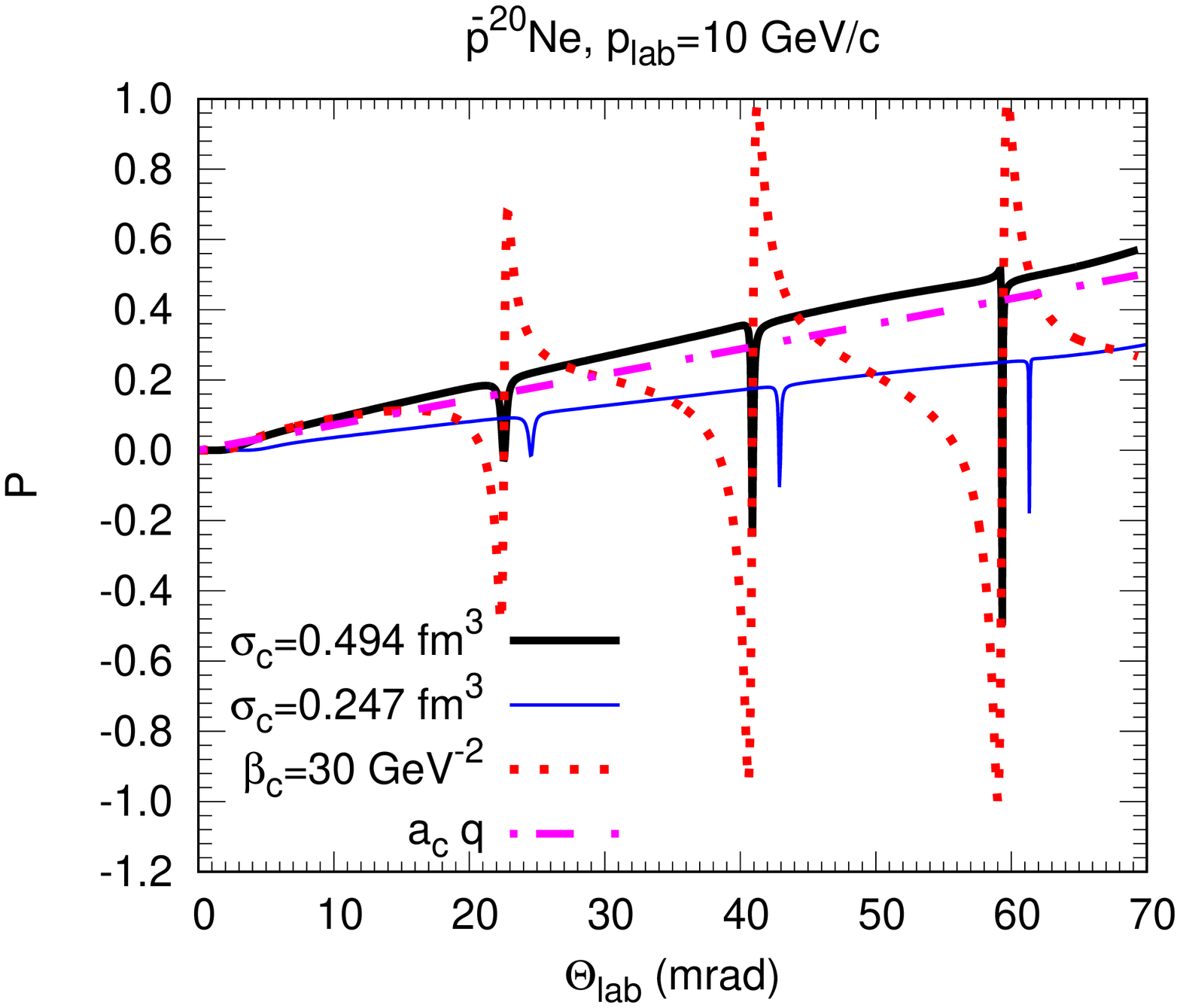}
\end{center}
\caption{\label{fig:pol_10gevc} Polarization as a function of the laboratory scattering angle for $\bar p$ elastic scattering on $^{20}$Ne
  at $p_{\rm lab}=10$ GeV/c. The meaning of lines is the same as in Fig.~\ref{fig:pol_LEAR}.
  For better visibility of the diffractive structures the dependence $P(\Theta_{\rm lab})$ for the smaller value of $\sigma_c$ (thin solid line) 
  is shifted to the right by 2 mrad.}
\end{figure}
The polarization for the $\bar p\,^{20}$Ne elastic scattering at 10 GeV/c is displayed in Fig.~\ref{fig:pol_10gevc}.
The values of the strength of the spin-orbit interaction have been calculated from Eq.(\ref{sigma_c_plab}) by assuming
that $\sigma_c(0.608~\mbox{GeV/c})=1.4$ (default value) and $0.7$ fm$^3$.
In calculations with the momentum slope of the spin-orbit interaction, $\beta_c=12.7$ GeV$^{-2}$
(default value at $p_{\rm lab}=10$ GeV/c), apart from narrow discontinuities at the diffractive minima, the polarization grows
almost linearly with scattering angle and reaches values of about $0.3\div0.6$ already at $\Theta_{\rm lab}=70$ mrad.
On the other hand, in calculation with a stiffer slope, $\beta_c=30$ GeV$^{-2}$, the polarization reveals
diffractive pole structures extended in $\Theta_{\rm lab}$. The large positive (negative) polarization is obtained
for the scattering angles slightly larger (smaller) than the diffractive minima.
This qualitative difference can be explained as follows. In the default calculation (with $\beta_c = \beta_p$)
the real and imaginary parts of the amplitudes $F$,$G$ change sign very close to the diffractive minima in $\Theta_{\rm lab}$.
Thus, the nonmonotonic structures in the polarization are strongly localized in $\Theta_{\rm lab}$ near diffractive minima in this case.
However, in the calculation with $\beta_c > \beta_p$ the zeros of the real and imaginary parts of the amplitude $G$ appear
at smaller $\Theta_{\rm lab}$ (i.e. at smaller momentum transfer). This causes a pronounced diffractive pattern in the polarization.
At the diffractive maxima ($\Theta_{\rm lab}=28,~47$ and $66$ mrad, see solid line in Fig.~\ref{fig:alpha_dep}) the polarization
experiences the monotonic growth with the scattering angle in all our calculations.
Thus, within uncertainty of the model parameters, the polarization in the angular regions near diffractive maxima
can be expressed by the area between the thick and thin solid lines in Fig.~\ref{fig:pol_10gevc}. 

\subsection{Antiproton absorption on nuclei} 
\label{results_Abs}

\begin{figure}
\begin{center}
   \includegraphics[scale = 0.40]{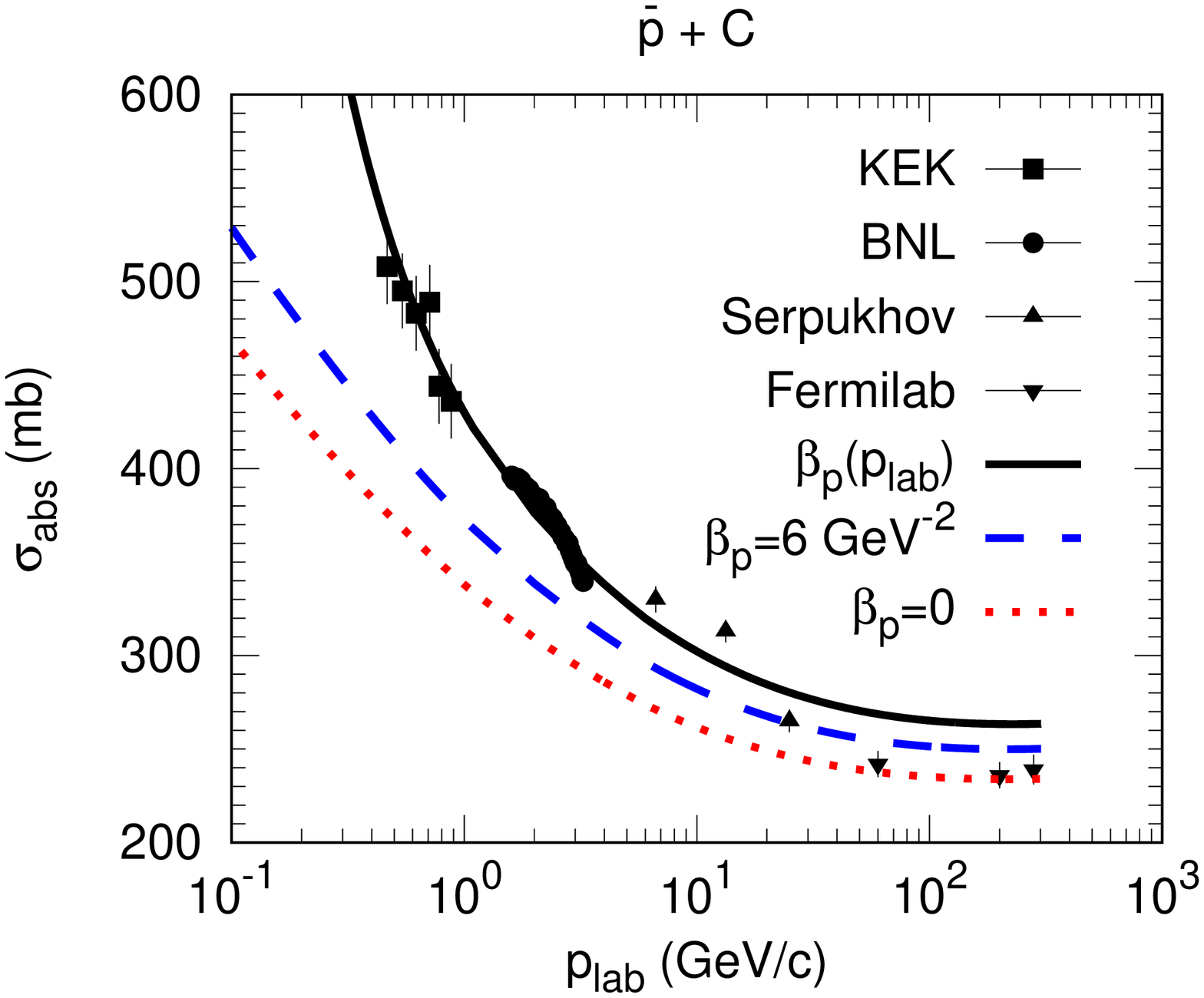}
   \includegraphics[scale = 0.40]{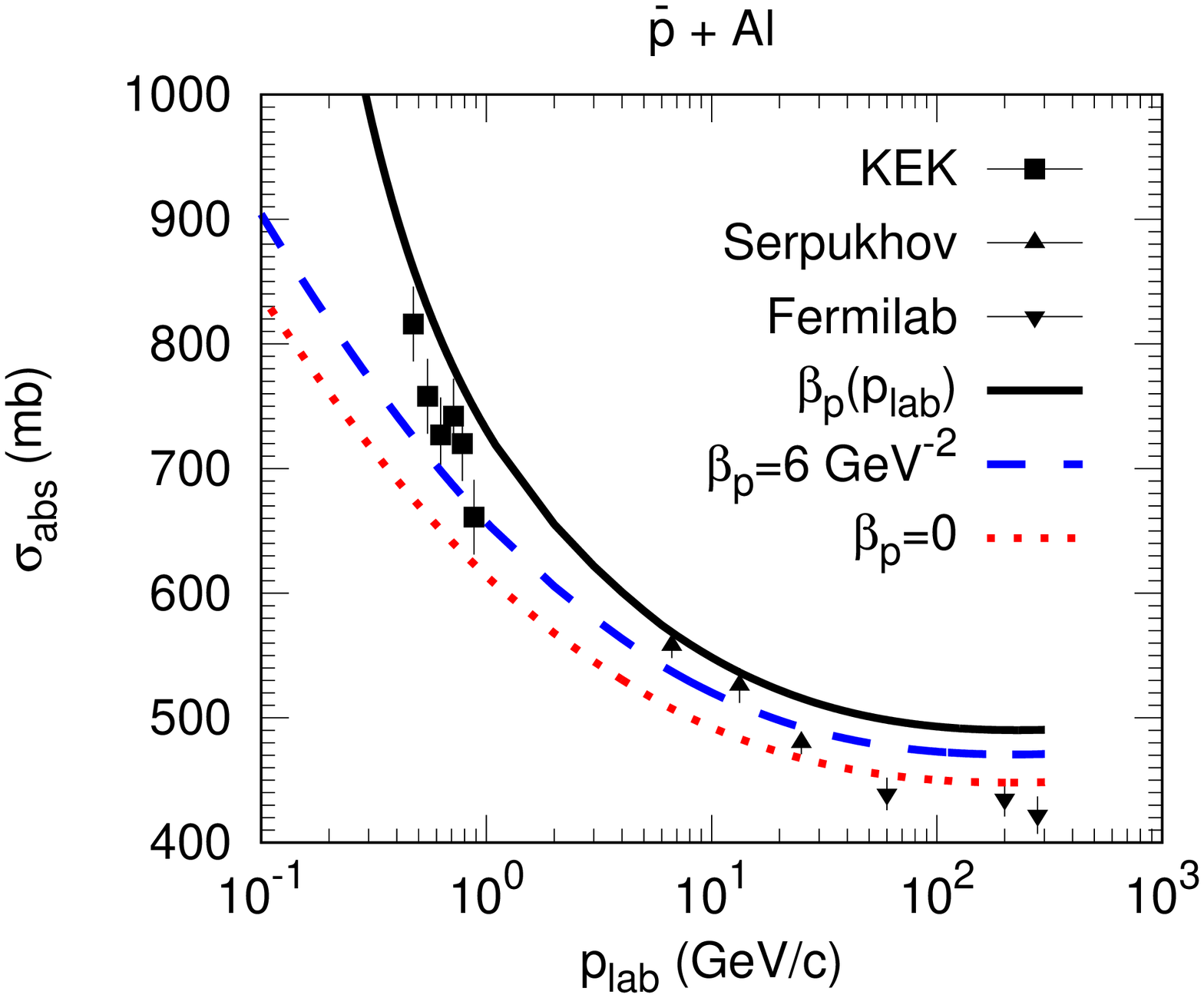}
   \includegraphics[scale = 0.40]{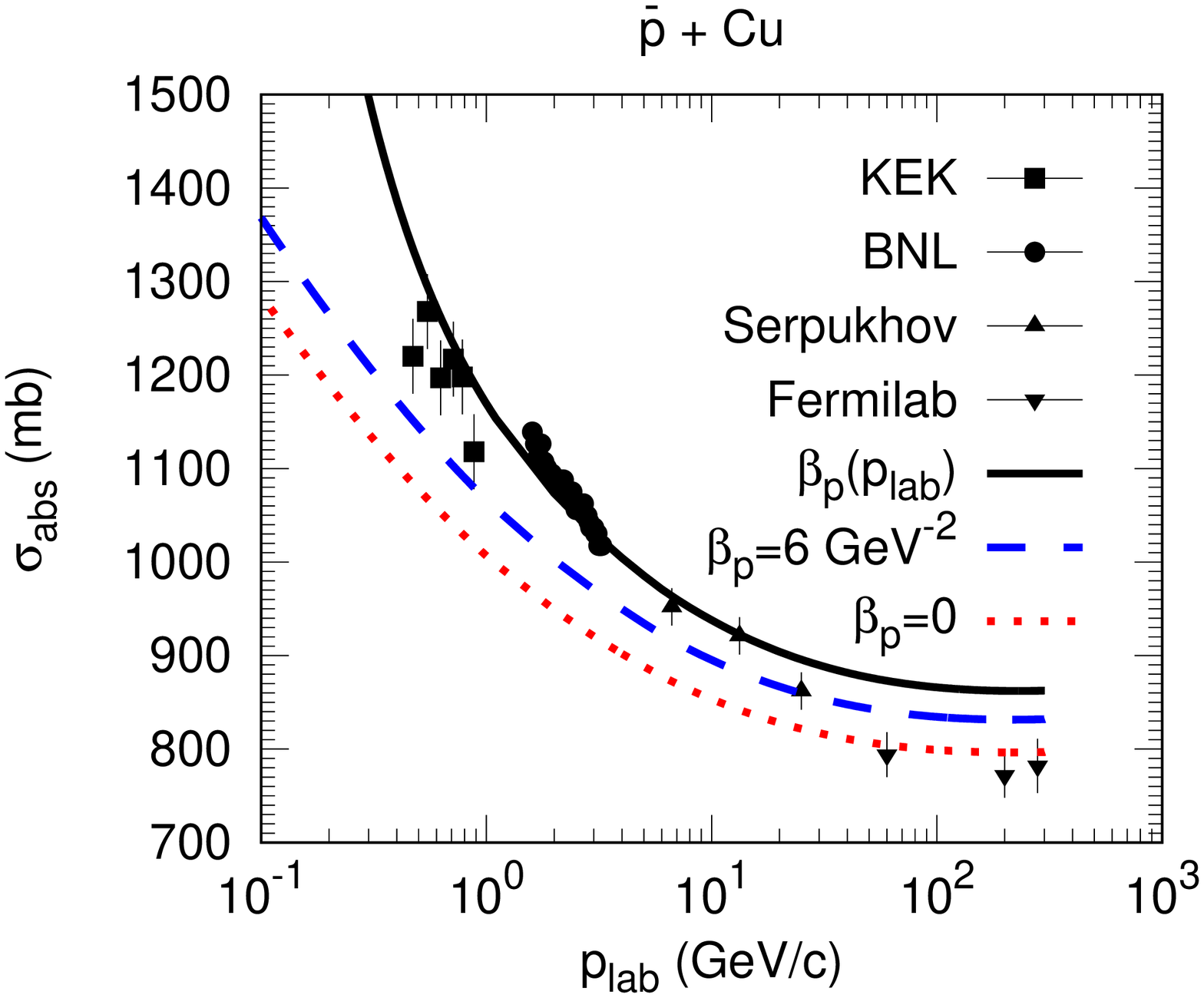}
\end{center}
\caption{\label{fig:abs} $\bar p$ absorption cross section on $^{12}$C, $^{27}$Al and $^{63}$Cu target nuclei as a function of beam momentum.
The calculation using slope parameter, $\beta_p$, of the momentum transfer dependence of the elastic $\bar p p$ amplitude in the parameterization
of ref. \cite{Kondratyuk:1986cq} is shown by solid line. The calculations with $\beta_p=6$ GeV$^{-2}$ and $\beta_p=0$ are shown, respectively,
by dashed and dotted lines. The experimental data are from ref. \cite{Nakamura:1984xw} (filled boxes), 
ref. \cite{Abrams:1972ab} (filled circles), ref. \cite{Denisov:1973zv} (filled triangles) and ref. \cite{Carroll:1978hc}
(filled upside-down triangles).}
\end{figure}
Comparison of our calculations with world data on $\bar p$ absorption cross section on carbon, aluminum and copper targets is shown 
in Fig.~\ref{fig:abs}.  The calculation using Eqs.(\ref{sigma_abs}),(\ref{chi_N_central})
with momentum-transfer-dependent slopes is in a good agreement with experimental data at $p_{\rm lab} \ltsim 10$ GeV/c (except for the KEK
data point at $p_{\rm lab}=0.881$ GeV/c for Al and Cu). However, this calculation overpredicts experimental absorption cross sections
at $p_{\rm lab} \sim 60\div300$ GeV/c by $\sim 15\%$. This overprediction is not unexpected, since with increasing beam momentum
above $\sim 10$ GeV/c the Glauber theory gradually looses its applicability due to increasing importance of the inelastic intermediate
states \footnote{According to the Gribov formalism (see \cite{Frankfurt:2011cs} and refs. therein), the inelastic diffraction increases
  the shadowing correction to the total pion-deuteron cross section, i.e. reduces the total pion-deuteron cross section with respect
  to the Glauber formula above $p_{\rm lab}=6$ GeV/c.}.

It is in order to note here that several previous GM calculations of absorption cross sections (cf. \cite{Abrams:1972ab,Larionov:2009tc})
were done in the semiclassical approximation, Eq.(\ref{sigma_abs_class}), neglecting the momentum transfer dependence of the elementary amplitude,
although a more rigorous way is to apply Eqs.(\ref{sigma_abs}),(\ref{chi_N_central}) (cf. \cite{Lenske:2005nt}).
The folding formula, Eq.(\ref{chi_N_central}), for the phase shift function is a consequence of quantum mechanical treatment of the nonlocality
of the interaction. Hence, it is useful to quantify the error introduced when one
applies the semiclassical Eq.(\ref{sigma_abs_class}) for the antiproton absorption cross sections.
As we see from Fig.~\ref{fig:abs}, the semiclassical approximation ($\beta_p=0$, dotted lines)
gives substantially smaller absorption cross section, in-particular, at low beam momenta
\footnote{Adding Coulomb and nuclear potentials beyond eikonal approximation \cite{Larionov:2009tc} improves the agreement
  with experimental data.}.
For orientation, absorption cross sections with the slope of $6$ GeV$^{-2}$, i.e about two times less than the plateau value
of $\beta_p$ at $p_{\rm lab} \sim 10^2\div10^3$ GeV/c (see Fig.~\ref{fig:slopes} above), are also shown in Fig.~\ref{fig:abs}.
Thus, the $\bar p A$ absorption cross section is strongly sensitive to the slope of the momentum transfer dependence
of the $\bar p N$ elastic amplitude. 
  
\begin{figure}
\begin{center}
  \includegraphics[scale = 0.40]{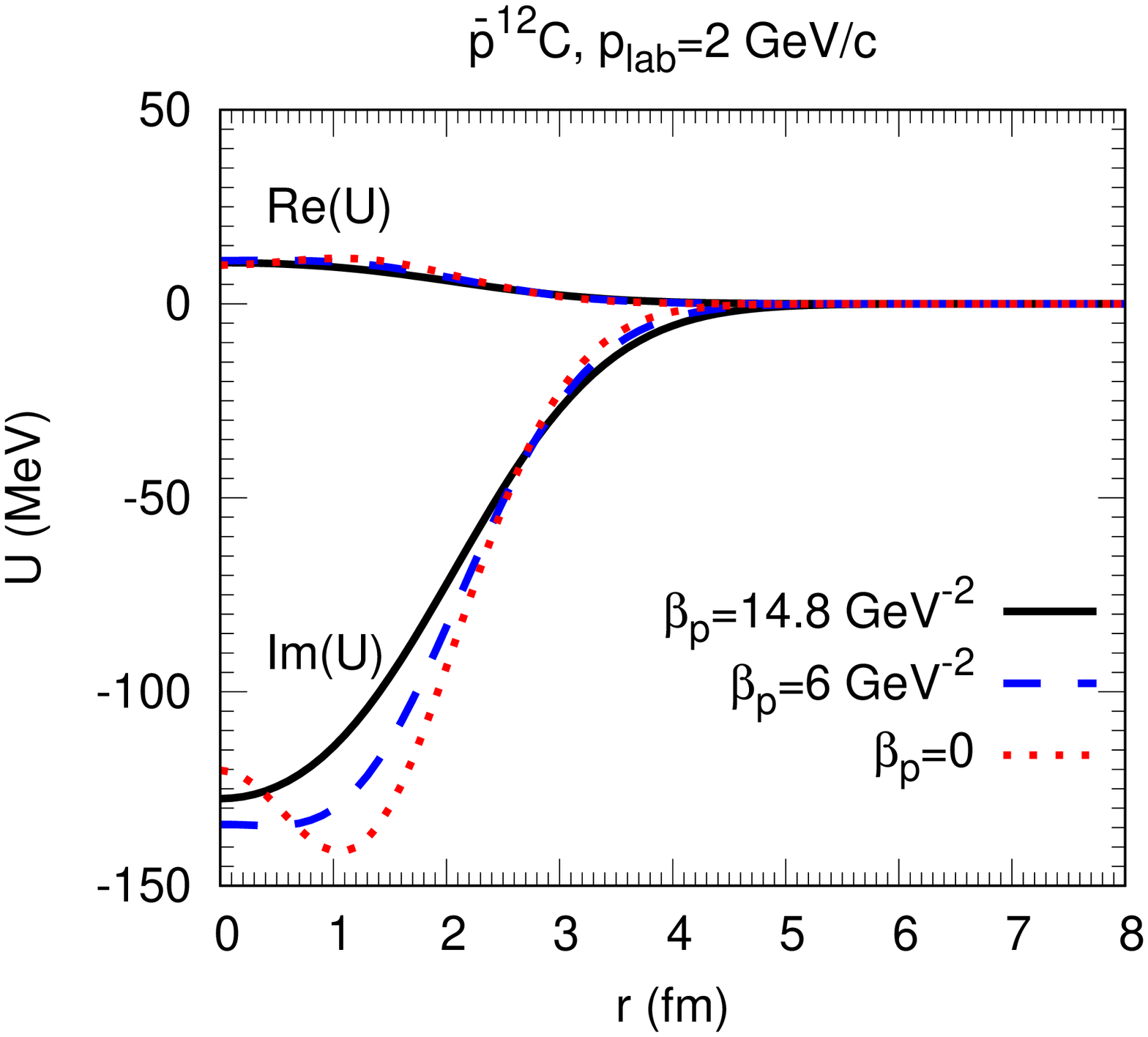}
  \includegraphics[scale = 0.40]{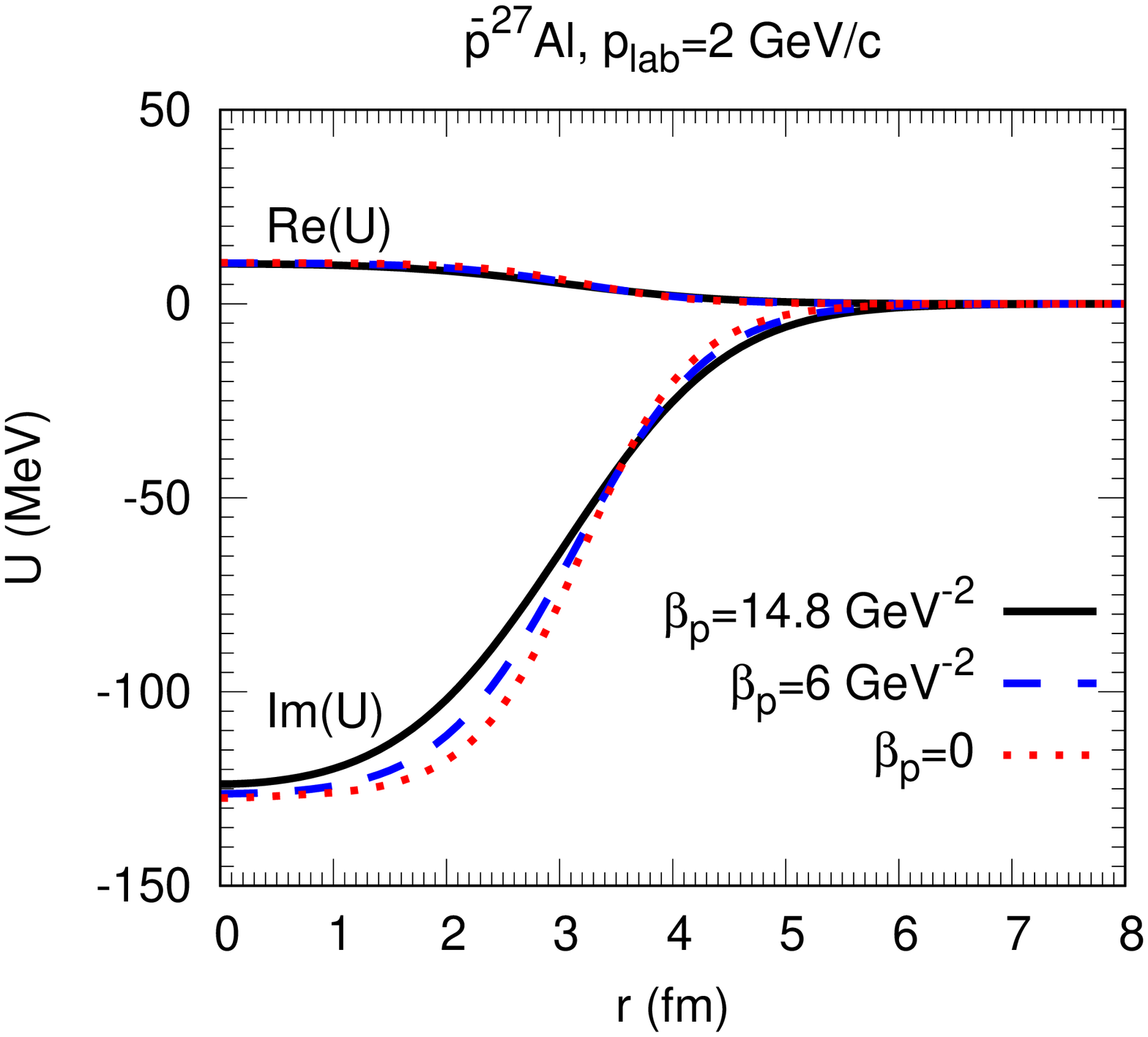}
  \includegraphics[scale = 0.40]{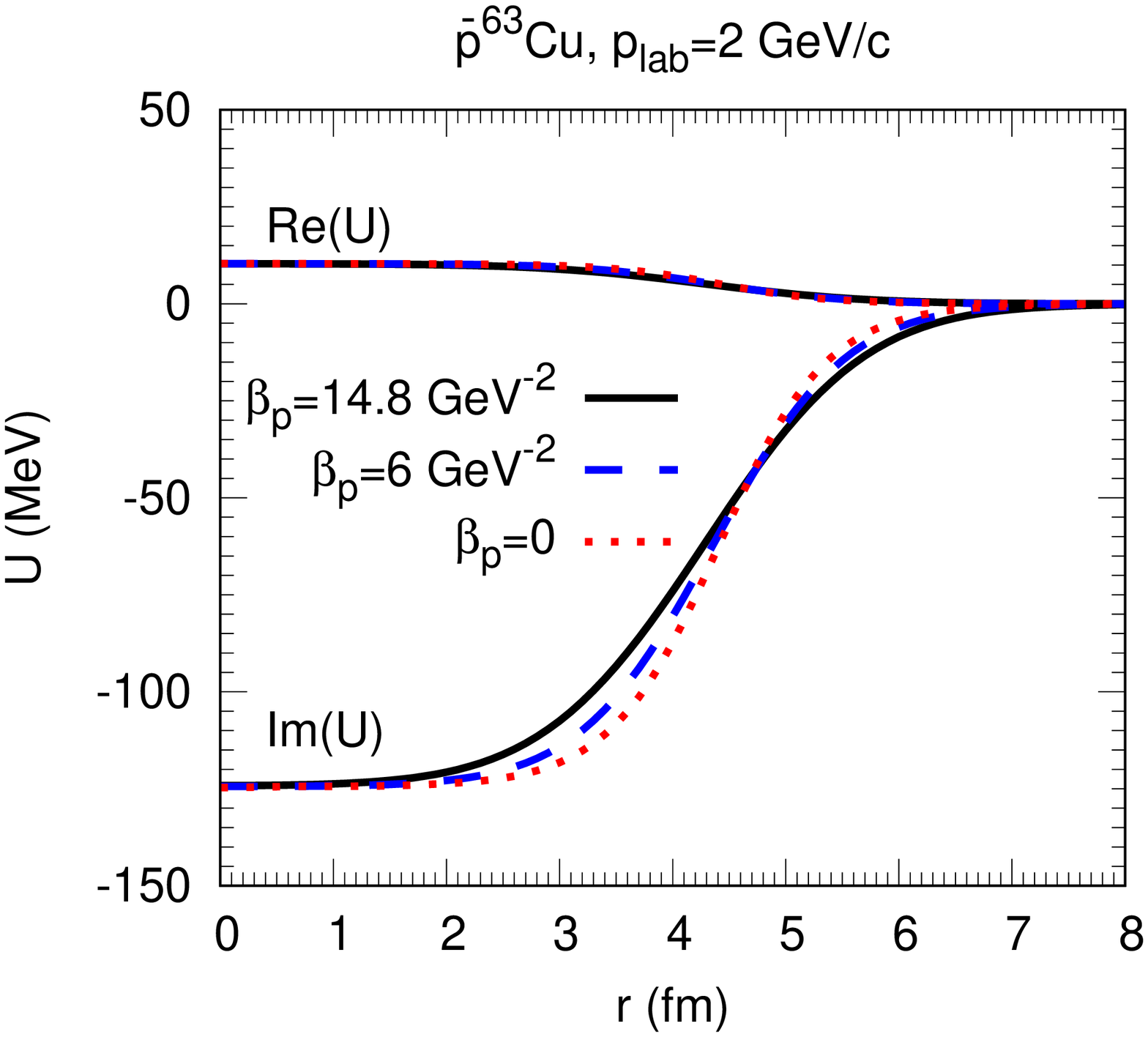}  
\end{center}
\caption{\label{fig:U} The real and imaginary parts of $\bar p$ nuclear central optical potential
at $p_{\rm lab}=2$ GeV/c for $^{12}$C, $^{27}$Al and $^{63}$Cu nuclei. Line notations are the same as in Fig.~\ref{fig:abs}.}
\end{figure}
This sensitivity can be explained in terms of the antiproton optical potential, Eq.(\ref{U}),
which is plotted in Fig.~\ref{fig:U}. 
A finite slope cuts the nuclear form factor at large momentum transfer
which results in a more smooth and extended in space optical potential.
According to Eq.(\ref{sigma_abs_U}) the absorption cross section rises for such extended potentials.

It is interesting that according to empirical data \cite{Agashe:2014kda} at high beam momenta
the antiproton-nucleus potential becomes weakly repulsive since $\alpha_p < 0$.
This is in contrast to the predictions of the relativistic mean field models based on $G$-parity transformation,
cf. refs. \cite{Teis:1994ie,Larionov:2009tc,Gaitanos:2011ej}. This difference requires further studies.

\section{Summary and conclusions}
\label{concl}

We have performed the Glauber model calculations of antiproton-nucleus scattering and absorption
at the beam momenta of $\sim 0.5\div10$ GeV/c. The $\bar p N$ scattering amplitude used in the model includes
central and spin-orbit interactions. The latter has been chosen to reproduce the $\bar p$ polarization
in elastic scattering on carbon measured at LEAR. We have also taken into account the Coulomb interaction and recoil
corrections and included realistic proton and neutron density profiles both for light and heavy nuclei.
The model successfully describes the angular differential cross sections of $\bar p A$ elastic scattering
at 0.608 GeV/c measured at LEAR and $pA$ elastic+quasielastic scattering at 9.92 GeV/c and 19.85 GeV/c measured
at BNL, in-line with the earlier theoretical analyses. This allows us to predict $\bar pA$ angular differential
elastic and quasielastic scattering cross sections and polarization at the beam momenta $\sim 10$ GeV/c
which can be summarized as follows:
\begin{itemize}

\item Angular differential cross sections of antiproton elastic scattering
  on nuclei  have significantly deeper diffractive minima as compared
  to the proton-nucleus elastic cross sections.

\item The depths of the diffractive minima are strongly sensitive to the ratios of
      the real-to-imaginary parts of the antiproton-proton and antiproton-neutron elastic amplitudes.

\item Strong polarization, $P = 0.3\div0.6$ is expected at forward scattering angles, $\Theta_{\rm lab} \simeq 70$ mrad.

\end{itemize}

We have also calculated the antiproton absorption cross sections on nuclei and compared our results with available 
experimental data at $p_{\rm lab} = 0.5\div300$ GeV/c.  The Glauber model describes low energy (KEK) data
at $p_{\rm lab} = 0.5\div0.9$ GeV/c
surprisingly well and seems to not leave room for non-eikonal effects caused by attractive Coulomb
and nuclear potentials provided the momentum transfer dependence is taken into account in the $\bar p N$ amplitude
\footnote{According to refs. \cite{Lee:2013rxa,Lee:2015hma}, the non-eikonal effects are very important at even lower beam momenta.}.
At high beam momenta, $p_{\rm lab} =60\div300$ GeV/c, we have figured out, however, that the usual semiclassical
absorption formula describes the Fermilab data in the best way, although the Glauber theory becomes hardly applicable here due to
missing inelastic shadowing corrections \cite{Frankfurt:2011cs}.

The eikonal approximation significantly simplifies calculations not only for elastic scattering and absorption
discussed here, but also for more complex processes involving production of new particles. This is because the calculations
based on eikonal approximation contain only (multiple) folding integrals and do not require iterations, like, 
e.g. in DWBA technique. Successful application of the Glauber model to the $\bar p A$ elastic scattering 
and absorption in this work is a convincing argument to apply eikonal approximation, in particular, for the initial 
and final state interactions in the exclusive hypernuclei production in $\bar p A$ reactions at FAIR energies. 
This study is currently in progress.

To conclude, the measurements of the $\bar pA$ elastic scattering, polarization and absorption at the
beam momenta $\sim 2\div10$ GeV/c would provide new important information on $\bar p N$ elastic amplitude at small
momentum transfers. This may potentially lead to tighter constraints on the parameters of the Regge model
of high-energy $\bar p p$ and $\bar p n$ scattering. Such studies can be performed in the forthcoming
PANDA experiment at FAIR. 

\begin{acknowledgments}
  The authors are grateful to Ulrich Mosel for the interest to this work and for highlighting
  the difference with the relativistic mean field predictions for $\mbox{Re}(U)$.
  This work was supported by the Deutsche Forschungsgemeinschaft (DFG) under Grant No. Le439/9.
\end{acknowledgments}

\bibliography{pbarEl}

\appendix

\section{Proton and neutron densities}
\label{SPdensities}

Let us consider the case when the single-particle state $j$ is characterized by the set of quantum numbers $\{nlJJ_z\tau\}$,
where $n$ is the principal quantum number, $l$ is the orbital angular momentum, $J$ and $J_z$ are the total angular momentum
and its $z$-component, respectively, and $\tau$ is the isospin quantum number. The single-particle wave function can be represented as
\begin{equation}
  \phi_j(x) \equiv \phi_{nlJJ_z\tau}(\bm{r},\lambda,T) = R_{nlJ}^{(\tau)}(r)\, \xi_\tau(T)
  \sum_{l_z=-l}^l  \sum_{\sigma=\pm1/2}  Y_{ll_z}(\hat{\bm{r}})\, \chi_\sigma(\lambda)\, \langle l l_z; \frac{1}{2} \sigma| J J_z \rangle~,  \label{phi_j_expl}
\end{equation}
where $R_{nlJ}^{(\tau)}(r)$ is the radial wave function, and $\chi_\sigma(\lambda)$ and $\xi_\tau(T)$ are, respectively,
the spin and isospin Pauli spinors. If there is no degeneracy of the energy levels with different principal quantum numbers,
the set of the radial wave functions is orthogonal:
\begin{equation}
  \int\limits_0^{+\infty} dr r^2 R_{n^\prime lJ}^{(\tau)*}(r) R_{nlJ}^{(\tau)}(r) = \delta_{n n^\prime}~. \label{R_orthog}
\end{equation}
The  Pauli spinors satisfy the orthogonality relations,
\begin{eqnarray}
  && \chi_{\sigma^\prime}^\dag \chi_\sigma = \sum_{\lambda=\pm1/2} \chi_{\sigma^\prime}^*(\lambda) \chi_\sigma(\lambda) = \delta_{\sigma\sigma^\prime}~,  \label{chi_orthog}   \\
  && \xi_{\tau^\prime}^\dag \xi_\tau = \sum_{T=\pm1/2} \xi_{\tau^\prime}^*(T) \xi_\tau(T) = \delta_{\tau\tau^\prime}~.   \label{f_orthog}
\end{eqnarray}
Using Eqs.(\ref{R_orthog})-(\ref{f_orthog}) one can prove the global orthogonality relation for the single-particle wave functions,
\begin{equation}
  \int d^3r \sum_{\lambda,T} \phi_{n^\prime l^\prime J^\prime J_z^\prime \tau^\prime}^*(\bm{r},\lambda,T)  \phi_{nlJJ_z\tau}(\bm{r},\lambda,T)
  = \delta_{nn^\prime} \delta_{ll^\prime} \delta_{JJ^\prime} \delta_{J_zJ_z^\prime} \delta_{\tau\tau^\prime}~,      \label{Glob_orhthog_expl}
\end{equation}
which is an explicit form of Eq.(\ref{Glob_orthog}). After simple transformations we come to the following expression
for the single-particle density (see Eq.(\ref{rho_j})):
\begin{equation}
  \rho_j(\bm{r}) \equiv \rho_{nlJJ_z\tau}(\bm{r}) = \sum_{\lambda,T} |\phi_{nlJJ_z\tau}(\bm{r},\lambda,T)|^2 = |R_{nlJ}^{(\tau)}(r)|^2
  \sum_{l_z,\sigma} |Y_{ll_z}(\hat{\bm{r}})|^2\, |\langle l l_z; \frac{1}{2} \sigma| J J_z \rangle|^2~.  \label{rho_j_expl}
\end{equation}
If the radial wave function does not depend on the total angular momentum, i.e. when the spin-orbit interaction is neglected,
we can also obtain the expression for the densities summed over all possible states of the $(n,l,\tau)$-shell:
\begin{equation}
  \sum_{J=|l-1/2|}^{l+1/2} \sum_{J_z=-J}^J \rho_{nlJJ_z\tau}(\bm{r}) = |R_{nl}^{(\tau)}(r)|^2 \sum_{l_z,\sigma}  |Y_{ll_z}(\hat{\bm{r}})|^2
     \sum_{JJ_z} |\langle l l_z; \frac{1}{2} \sigma| J J_z \rangle|^2
     = \frac{2(2l+1)}{4\pi} |R_{nl}^{(\tau)}(r)|^2~,     \label{l_shell}
\end{equation}
where we used the orthogonality relation for the Clebsch-Gordan coefficients and the addition theorem for the spherical harmonics.
As expected, Eq.(\ref{l_shell}) integrated over $d^3r$ gives the maximal occupation number, $2(2l+1)$, of the shell with
the orbital angular momentum $l$. 

The proton and neutron densities, $\rho_p(r)$ and $\rho_n(r)$, are calculated by summing the corresponding single-particle densities
as expressed by Eq.(\ref{rho_tau}). We have applied in calculations the Woods-Saxon parameterizations for $\rho_p(r)$ and $\rho_n(r)$
obtained by fitting the results of the Skyrme-Hartree-Fock calculations for a large set of nuclei. 

\end{document}